\documentclass[11pt]{article}
\usepackage[numbers,round,sort&compress]{natbib}
\usepackage{float}
\usepackage{authblk}
\usepackage{geometry}
\usepackage{indentfirst}
\usepackage[bf,footnotesize]{caption}
\usepackage[colorlinks=true,allcolors=blue]{hyperref}
\usepackage[english]{babel}
\usepackage{pdfpages}
\usepackage{amsmath}
\usepackage{amssymb}
\usepackage{amsthm} 
\usepackage{bm}
\usepackage{booktabs}
\usepackage{changepage}
\usepackage{color}
\usepackage{eucal}
\usepackage{fullpage}
\usepackage{graphicx}
\usepackage{hhline}
\usepackage{lineno}
\usepackage{makecell}
\usepackage[sc]{mathpazo}
\usepackage{mathrsfs}
\usepackage{mathtools}
\usepackage{multirow}
\usepackage{setspace}
\usepackage{stmaryrd}
\usepackage[noend]{algpseudocode}
\usepackage{algorithmicx,algorithm}
\usepackage{bbm}

\newcommand*\patchAmsMathEnvironmentForLineno[1]{%
    \expandafter\let\csname old#1\expandafter\endcsname\csname #1\endcsname
    \expandafter\let\csname oldend#1\expandafter\endcsname\csname end#1\endcsname
    \renewenvironment{#1}%
    {\linenomath\csname old#1\endcsname}%
    {\csname oldend#1\endcsname\endlinenomath}}%
    \newcommand*\patchBothAmsMathEnvironmentsForLineno[1]{%
    \patchAmsMathEnvironmentForLineno{#1}%
    \patchAmsMathEnvironmentForLineno{#1*}}%
    \AtBeginDocument{%
    \patchBothAmsMathEnvironmentsForLineno{equation}%
    \patchBothAmsMathEnvironmentsForLineno{align}%
    \patchBothAmsMathEnvironmentsForLineno{flalign}%
    \patchBothAmsMathEnvironmentsForLineno{alignat}%
    \patchBothAmsMathEnvironmentsForLineno{gather}%
    \patchBothAmsMathEnvironmentsForLineno{multline}%
}

\definecolor{rv}{rgb}{0, 0, 1}
\definecolor{sgreen}{RGB}{0, 106, 20}
\definecolor{highlight}{rgb}{1,0,0}

\newcommand\qc{q_{\left(n_{C},n_{D}\right)\mid C}}
\newcommand\qd{q_{\left(n_{C},n_{D}\right)\mid D}}
\newcommand\summ{\sum_{n_{C}+n_{D}=\ell-1}}
\newcommand\bcratio{\left( \frac{b}{c} \right)^{\ast}}

\title{Hyperedge approximation for stochastic processes on higher-order networks}

\author
{Anzhi Sheng$^{1,2}$, Alex McAvoy$^{3,4}$, Ye Tian$^{1}$, Silun Zhang$^{2}$, Angela Fontan$^{1}$, Joshua B. Plotkin$^{5,6}$\\
\footnotesize{$^{1}$Department of Decision and Control Systems, KTH Royal Institute of Technology, Stockholm 10044, Sweden\\
$^{2}$Department of Mathematics, KTH Royal Institute of Technology, Stockholm 10044, Sweden\\
$^{3}$School of Data Science and Society, University of North Carolina at Chapel Hill, Chapel Hill, NC 27599, USA \\
$^{4}$Department of Mathematics, University of North Carolina at Chapel Hill, Chapel Hill, NC 27599, USA \\
$^{5}$Department of Biology, University of Pennsylvania, Philadelphia, PA 19104, USA\\
$^{6}$Center for Mathematical Biology, University of Pennsylvania, Philadelphia, PA 19014, USA}\\
}

\date{}

\begin{document}
\maketitle

\begin{abstract}
\noindent Graphs are a standard framework for describing dynamical processes shaped by pairwise interactions among agents. But many systems involve interactions in groups of three or more agents.
Here, we develop a method of "$\ell$-hyperedge approximation", a framework to analyze stochastic population processes on regular hypergraphs, in which each individual belongs to $k$ groups of size $\ell$. The framework accommodates both higher-order interactions that determine payoffs and higher-order processes for updating states in response to payoffs. Applied to evolutionary game dynamics, the framework generalizes the classical pairwise result on benefits and costs, $b/c>k$, that favors the spread of cooperation; and it provides critical benefit-to-cost ratios for nonlinear $\ell$-player public goods games that cannot be reduced to pairwise interactions. Applied to complex contagions, where inheritance of states occurs within hyperedges rather than along parent-offspring edges, the framework gives a closed-form result for the fixation probability, which shows how a complexity parameter governs the spread of rare types. Coupling the two processes produces a single stochastic model of payoff-biased complex contagion in structured populations. These results extend pair approximation from graphs to hypergraphs, accommodating multi-way interactions and inheritance structures with no pairwise analog.
\end{abstract}

\section{Introduction}
Graphs provide a useful framework to describe the microscopic structure of complex systems \cite{albert2002statistical,newman2003structure,barrat2004architecture,radicchi2013abrupt}.
Each unit or agent is represented as a node, and interactions are captured by edges between pairs of nodes.
Pair approximation has become a standard and powerful technique for analyzing stochastic dynamics on graphs, with applications to the Ising model in physics \cite{dorogovtsev2002ising}, social dilemmas in evolutionary game theory \cite{ohtsuki:simple:2006,su2022evolution}, and predator--prey dynamics in ecology \cite{durrett1994importance}.
The method tracks the statistical properties of neighboring pairs while averaging over finer details of the network.

But many real interactions involve more than two individuals at once \cite{lambiotte2019networks,battiston2021physics,grilli2017higher,levine2017beyond,fontan2025collective}.
Scientific collaborations, public-goods contributions, and group discussions all entail simultaneous interaction among several agents, with outcomes that pairwise decomposition cannot recover \cite{battiston2020networks}.
Such higher-order interactions in a structured population are naturally described by hypergraphs, in which a hyperedge connects an arbitrary number of nodes.
Two canonical group games illustrate the range: the public goods game, in which each member's contribution benefits all other players \cite{santos:social:2008,li2016evolutionary,alvarez2021evolutionary}, and the multi-player donation game, in which a single cooperator confers benefits to the entire group \cite{rand2014static,mcavoy2020social,sheng2023evolutionary}.

Higher-order interactions can also shape how states are updated over time, independent of gameplay itself.
On pairwise graphs, imitation dynamics propagate from parent to offspring along an edge
--- simple contagion \cite{mobilia2003does,durrett2005can,santos2005scale,castellano2009statistical,traulsen2010human,barrat2008dynamical,pinheiro2014origin,iacopini2018network}.
On hypergraphs, updates occur at the group level, and the adoption rate may depend nonlinearly on the number of neighbors already holding a given state (so-called ``complex'' contagion) \cite{centola2007complex,centola2010spread,campbell2013complex,karsai2014complex,centola2018behavior,vasconcelos2019consensus,chiba2024social}.
In the field of opinion dynamics, for example, the probability that an individual adopts a particular opinion may increase synergistically, or saturate, with the number of peers already holding it.
Such nonlinearities in the update process, even in the absence of fitness effects, fall outside the scope of pair approximation and of existing coalescent-based methods \cite{allen:evolutionary:2017,mcavoy2021fixation,su2023strategy,sheng2024strategy}.

Here we develop the $\ell$-hyperedge approximation, a framework for analyzing stochastic processes on regular hypergraphs in which every node is a member of $k$ hyperedges, each of which has order $\ell$.
The framework accommodates higher-order interactions both in the payoff-generating phase---evolutionary games on hypergraphs---and also in the state-updating phase---complex contagion.
We benchmark this analytic approximation against the fixation probability of a rare strategy or opinion \cite{nowak:2004:emergence}, and we extend it to a combined model in which payoff-biased selection and complex contagion act simultaneously.

For evolutionary games on hypergraphs, where higher-order interactions shape only the interaction and payoff-generating phase, analysis of the $\ell$-player donation game produces a result that generalizes the classical result for pairwise graphs \cite{ohtsuki:simple:2006}, namely cooperation is favored whenever $b/c > k$.
For the nonlinear $\ell$-player public goods game, we compute the critical benefit-to-cost ratio to favor cooperation as a function of the degree $k$ and a benefit factor $\delta$ that encodes synergy or antagonism among private contributions.
For neutral complex contagions where higher-order interactions shape only the inheritance process, we derive a closed-form expression for the fixation probability and show how the complexity parameter governs the rate at which a rare opinion spreads.
Coupling the two processes yields a single stochastic model of payoff-biased complex contagion, and we characterize how conformity and innovation shift the threshold for the spread of cooperation.

\section{Model}

\subsection{Stochastic processes on hypergraphs}

\begin{figure}[p]
\centering
\includegraphics[width=0.9\textwidth]{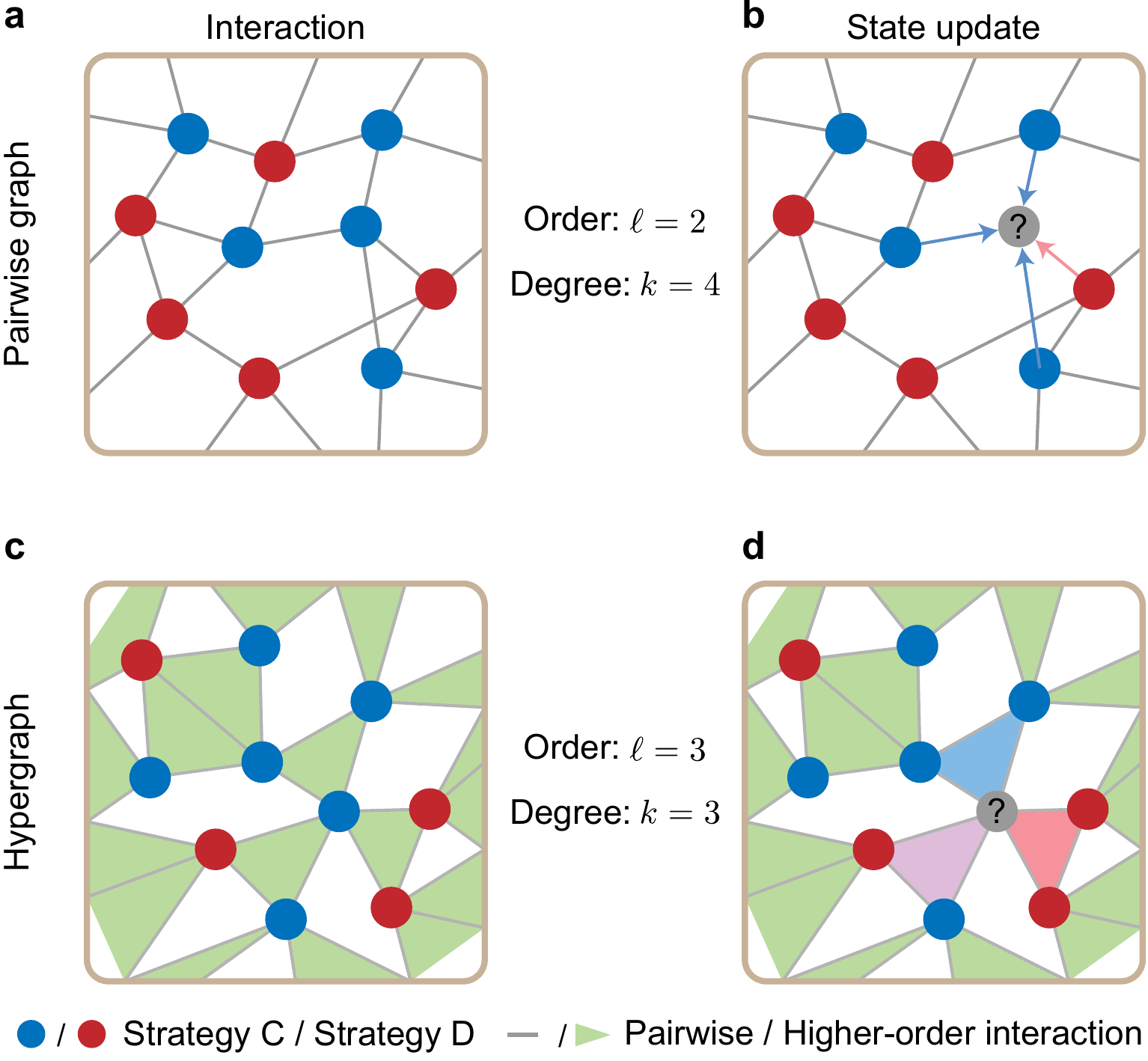}
\caption{\textbf{
Strategy evolution on graphs and hypergraphs.} The population structure is represented by a regular network of order $\ell$, we consider $\ell =2$ (pairwise graph) and $\ell =3$ (third-order hypergraph). 
Interactions (gameplay) are represented by gray lines and green triangles, respectively.
\textbf{a,c}, Each individual occupies a node and adopts either strategy $C$ (blue circles) or strategy $D$ (red circles). Each individual then interacts with their neighbors in either two- or three-player games, and receives accumulated payoffs.
\textbf{b,d}, At each time step, a random individual $i$ (labeled ``?'') is selected uniformly to update their strategy.
The probability of adopting strategy $C$ is positively related to: (i) the number of (hyper)edges containing $n_{C}$ neighbors in state $C$ and $n_{D}$ neighbors in state $D$, and (ii) the rate of adopting $C$ in such edges, denoted by $F_{C}\left(n_{C},n_{D}\right)$  if $i$ is in state $D$, or by $G_{C}\left(n_{C},n_{D}\right)$ if $i$ is in state $C$. Corresponding expressions provide the chance of adopting strategy $D$.
On a pairwise graph, strategy updating reduces to standard death-birth dynamics \cite{ohtsuki:simple:2006} (\textbf{b}), where the selected individual $i$ simply imitates the strategy of a neighbor with a probability that depends on their payoffs.
}
\label{fig:illustration}
\end{figure}

We describe a broad class of stochastic processes on hypergraphs, including those with frequency-dependent selection.
Consider a population of $N$ individuals whose structure is represented by a regular hypergraph of order $\ell$: each node denotes an individual, and each hyperedge captures an interaction within a group of $\ell$ individuals.
Figures~\ref{fig:illustration}a,c illustrate two cases we analyze in detail: $\ell = 2$ (pairwise graph) and $\ell = 3$ (third-order hypergraph).

Each individual holds one of two strategies, $C$ or $D$, and participates in $k$ order-$\ell$ interactions. The hyperedges attached to a focal individual are classified into $2\ell$ categories of the form $(X, n_C, n_D)$, where $X \in \{C, D\}$ is the strategy of the focal individual, and $n_C + n_D = \ell - 1$ is the composition of the $\ell - 1$ remaining neighbors in the hyperedge.
Let $k_{n_C, n_D}$ denote the number of hyperedges of composition $(X, n_C, n_D)$ attached to the focal individual, so that $\summ k_{n_C, n_D} = k$.

After all interactions occur, a random individual $i$ is selected to update their strategy.
If $i$ previously adopted $D$, the probability that $i$ updates their strategy to $C$ is
\begin{equation}
    p_{D\to C} = \frac{\summ  k_{n_{C},n_{D}} F_{C}\left(n_{C},n_{D};\theta\right)}{\summ k_{n_{C},n_{D}} \left(F_{C}\left(n_{C},n_{D};\theta\right) + F_{D}\left(n_{C},n_{D};\theta\right) \right)},
\end{equation}
where $F_{C}\left(n_{C},n_{D};\theta\right)$ and $F_{D}\left(n_{C},n_{D};\theta\right)$ denote the weights associated to the chances of $i$ adopting $C$ and $D$ in a hyperedge of type $\left(D,n_{C},n_{D}\right)$, respectively.
Likewise, if $i$ previously adopted $C$, the probability of remaining at $C$ is
\begin{equation}
    p_{C\to C} = \frac{\summ  k_{n_{C},n_{D}} G_{C}\left(n_{C},n_{D};\theta\right)}{\summ k_{n_{C},n_{D}} \left(G_{C}\left(n_{C},n_{D};\theta\right) + G_{D}\left(n_{C},n_{D};\theta\right) \right)},
\end{equation}
where $G_{C}\left(n_{C},n_{D};\theta\right)$ or $G_{D}\left(n_{C},n_{D};\theta\right)$ denote the weights associated to the chances of $i$ adopting $C$ or $D$ in a hyperedge of type $\left(C,n_{C},n_{D}\right)$.

The parameter $\theta \ge 0$ is the intensity of selection \cite{ohta2002near}.
We focus on the regime of weak selection, $\theta \ll 1$, a standard assumption in evolutionary biology and sociophysics \cite{ohtsuki:simple:2006,wu2010universality,akashi1995inferring,allen:evolutionary:2017,traulsen2005coevolutionary,szabo2007evolutionary}.
We assume that, in the absence of selection ($\theta = 0$), the strategy dynamics are fair in expectation in the sense that no strategy is favored and the global frequency $p_C$ evolves as a martingale.

On a pairwise graph ($\ell = 2$), this update reduces to standard death-birth dynamics \cite{ohtsuki:simple:2006}: the focal individual copies the strategy of a single neighbor (Fig.~\ref{fig:illustration}b), perhaps biased by their relative payoffs.
On a hypergraph ($\ell \ge 3$), the focal individual is influenced simultaneously by all $\ell - 1$ co-players in the selected hyperedge, and the adoption rate may depend nonlinearly on their composition (Fig.~\ref{fig:illustration}d).
The hypergraph update is not reducible to an iteration of pairwise imitations, nor is there a single ``parent" from which a focal individual inherits their strategy.

\subsection{The $\ell$-hyperedge approximation and fixation probability}
To analyze the stochastic process above, we extend the classical pair approximation for regular graphs to an $\ell$-hyperedge approximation.
Let $p_{C}$ denote the global frequency of strategy $C$, and let $q_{\left(n_{C},n_{D}\right)\mid C}$ and $q_{\left(n_{C},n_{D}\right)\mid D}$ denote the conditional probabilities that, given a focal individual adopts $C$ or $D$, a randomly chosen hyperedge containing the focal individual has composition $(n_{C}, n_{D})$ among its $\ell - 1$ remaining positions.
Together, $p_{C}$ and the conditional frequencies determine the state of the system at the hyperedge level.

Under weak selection, the hyperedge frequencies $\qc$ and $\qd$ equilibrate on a faster timescale than the global frequency $p_{C}$.
At quasi-equilibrium, each conditional frequency can be expressed as a polynomial in $p_{C}$ (see Eq.~\ref{eq:qcqd} in Methods and SI Section 1.3).
A central identity governs these quasi-equilibrium relations:
\begin{equation}
    \summ n_{C} \left(\qc-\qd \right) = \frac{1}{k-1},
    \label{eq:identity}
\end{equation}
which indicates that a $C$-individual has, on average, one more $C$-neighbor than a $D$-individual across its $k-1$ other hyperedges.
Equation~\ref{eq:identity} generalizes the positive-assortment identity of Ohtsuki et al.~\cite{ohtsuki:simple:2006} to hypergraphs of arbitrary order.

Our primary object of interest is the fixation probability $\rho_{C}$: the probability that a single $C$ mutant, introduced at a uniformly random node in an otherwise all-$D$ population, eventually takes over the whole population.
Because the process has no mutation, the population eventually reaches one of two absorbing states (all-$C$ or all-$D$); $\rho_{C}$ quantifies how readily $C$ invades.

To derive the fixation probability $\rho_{C}$, we track the dynamics of $p_{C}$, $q_{\left(n_{C},n_{D}\right)\mid C}$, and $q_{\left(n_{C},n_{D}\right)\mid D}$ throughout the process using the $\ell$-hyperedge approximation (see SI Section 1).
Since $q_{\left(n_{C},n_{D}\right)\mid C}$ and $q_{\left(n_{C},n_{D}\right)\mid D}$ are polynomial functions of $p_{C}$, it is sufficient to track the expected change and the variance in $p_{C}$ over a single time step, denoted by $\mathbb{E}(\Delta p_{C})$ and $\text{Var}(\Delta p_{C})$, respectively.
Using the polynomial closure together with a diffusion approximation (see Eq.~\ref{eq:generalfixationprob} in Methods and SI Section 2), we obtain
\begin{align}
\rho_{C} &= \frac{1}{N} + \frac{\theta}{N} \sum_{m=0}^{\ell-1} \frac{1}{\left(m+1\right)\left(m+2\right)}\overline{\tau}^{\left(m\right)} + O\left(\theta^2\right) ,
\label{eq:generalfixationprobability}
\end{align}
for sufficiently large $N$, where the coefficients $\overline{\tau}^{\left(m\right)}$ are the coefficients of $2\mathbb{E}(\Delta p_{C})/\text{Var}(\Delta p_{C})$ when expanded as a polynomial in $p_{C}$ (see SI Section 2 for more details).

\section{Results}
We apply Eq.~\ref{eq:generalfixationprobability} to three classes of dynamics on hypergraphs: evolutionary games, where higher-order interactions determine payoffs; neutral complex contagion, where higher-order interactions determine state updates; and payoff-biased complex contagion, where both phases are affected.

\subsection{Evolutionary games on hypergraphs}
In the context of evolutionary game dynamics, strategy $C$ represents cooperation and strategy $D$ represents defection.
Each $\ell$th-order interaction is a game with $\ell$ concurrent players: a focal cooperator receives payoff $a_{n_{C}}$, and a focal defector receives payoff $b_{n_{C}}$ when $n_{C}$ of the $\ell-1$ co-players choose cooperation.

After all games have been played, each individual accumulates a payoff from their $k$ order-$\ell$ interactions. Then, a random individual $i$ is selected to update their strategy by imitating one of their neighbors.
Following the death-birth updating rule \cite{ohtsuki:simple:2006}, the probability that $i$ imitates a given neighbor $j$ is proportional to $j$'s fitness and to the number of interactions in which $i$ and $j$ both participate. Concretely,
\begin{equation}
\begin{aligned}
    F_{C}\left(n_{C},n_{D};\theta\right) &= n_{C} \exp\left({\theta u_{C}\left(n_{C},n_{D}\right)}\right),\ F_{D}\left(n_{C},n_{D};\theta\right) = n_{D} \exp\left( \theta u_{D}\left(n_{C},n_{D}\right)\right) ,\\
    G_{C}\left(n_{C},n_{D};\theta\right) &= n_{C} \exp\left(\theta v_{C}\left(n_{C},n_{D}\right)\right),\ G_{D}\left(n_{C},n_{D};\theta\right) = n_{D} \exp\left(\theta v_{D}\left(n_{C},n_{D}\right)\right),
    \label{eq:rate_game}
\end{aligned}
\end{equation}
where $u_{C}\left(n_{C},n_{D}\right)$ and $u_{D}\left(n_{C},n_{D}\right)$ represent the accumulated payoffs to a cooperator neighbor and a defector neighbor in the hyperedge $\left(D,n_{C},n_{D}\right)$, respectively, and $v_{C}\left(n_{C},n_{D}\right)$ and $v_{D}\left(n_{C},n_{D}\right)$ represent the corresponding payoffs to a cooperator and defector neighbor in the hyperedge $\left(C,n_{C},n_{D}\right)$.
Payoffs determine individuals' fitness $f=\exp(\theta u)$.
Substituting Eq.~\ref{eq:rate_game} into Eq.~\ref{eq:generalfixationprobability} gives the fixation probability of cooperation on hypergraphs of order $\ell$ (see Eq.~\ref{eq:prob_game}).

We focus on the cases $\ell = 2$ (standard graphs) and $\ell = 3$ (order-three hypergraphs).
We are interested in understanding when selection favors the evolution of cooperation, namely under what conditions the fixation probability of cooperation under weak selection exceeds that under neutral drift ($\rho_C > 1/N$).
For $\ell = 2$, this condition becomes
\begin{equation}
    (2 k^2-2k-1)a_0+(k^2+2k+1)a_1 > (2 k^2+k-1)b_0 +(k^2-k+1)b_1,
    \label{eq:l2}
\end{equation}
which recovers the well-known result derived by the pair approximation \citep{ohtsuki:simple:2006}.
When $\ell =3$, the condition becomes
\begin{equation}
\begin{aligned}
    &(12 k^3-16k^2+3k+3)a_0+(8 k^3-6k-6)a_1+(4 k^3+4k^2+3k+3)a_2\\
    &>(12 k^3-9k+3)b_0+(8 k^3 - 8k^2+6k-6) b_1+(4k^3 -4k^2 + 3k+3) b_2.
\end{aligned}
\label{eq:l3}
\end{equation}
A parallel condition for the relative success of cooperation over defection, $\rho_{C} > \rho_{D}$, is derived in Methods (Eq.~\ref{eq:relativesuccess}). 
This condition is identical to that for the absolute success ($\rho_C>1/N$, Eq.~\ref{eq:l3}) when the game is additive (i.e. linear), in which case a multi-player game on higher-order interactions can be reduced to a collective of pairwise games on pairwise edges.
However, when the game exhibits nonlinearity, higher-order interactions come into play, and the two conditions differ.

\subsubsection{Donation games}
Our first application of this general result is to determine what (if any)  multi-player payoff structure preserves the classical $b/c > k$ rule for the spread of cooperation.
Equation~\ref{eq:l3} reduces to $b/c > k$ under the assignment $a_0 = -c$, $a_1 = -c + b$, $a_2 = -c + 2b$, $b_0 = 0$, $b_1 = b$, $b_2 = 2b$.
More generally, we prove (see Eq.~\ref{eq:26} in Methods) that on hypergraphs of arbitrary order $\ell$, cooperation is favored whenever $b/c > k$, under the payoff scheme $a_{n_{C}} = -c + n_{C} b$ and $b_{n_{C}} = n_{C} b$.
We refer to this payoff structure as the \emph{$\ell$-player donation game}: a cooperator pays cost $c$ and provides benefit $b$ to each of its $\ell - 1$ co-players, while a defector pays nothing and provides nothing.
The total benefit provided by a single cooperator is $B = (\ell - 1) b$, so larger groups require a proportionally larger total benefit to sustain cooperation.

The intuition behind the $b/c>k$ rule on hypergraphs follows directly from the fundamental identity built into the $\ell$-hyperedge approximation (Eq.~\ref{eq:identity}).
From Eq.~\ref{eq:identity}, a cooperator has on average one additional cooperative neighbor, across its $k - 1$ other hyperedges, than a defector.
This assortment produces an expected payoff difference of $b - kc$ between a cooperator and a defector.
Cooperation is favored exactly when this difference is positive, which yields $b/c > k$ (see Supplementary Fig.~1).

The $\ell$-player donation game is linear and additive, and so its analysis reduces to a sequence of pairwise donation games (by rescaling cost to $c/(\ell - 1)$ and degree to $k(\ell - 1)$).
The theory of traditional pair approximation would therefore suffice to analyze the evolution of cooperation in this higher-order donation game; and so the result above is best seen as a consistency check that our method of higher-order analysis produces the expected, generalized result for higher-order donation games.
Furthermore, in this case, the condition for the absolute success of cooperation ($\rho_C > 1/N$) is equivalent to that for the relative success ($\rho_C > \rho_D$).

\subsubsection{Public goods games}
Our framework for studying higher-order interactions extends to non-linear multiplayer games, which cannot be studied by reduction to pairwise games.
As an illustrative example we compute the threshold for cooperation in the classic $\ell$-player public goods game, on a regular hypergraph.

In the public goods game, a cooperator pays a cost $c$ to contribute to a shared benefit that is distributed equally among all $\ell$ players (regardless of strategy), while a defector pays nothing.
The total benefit scales nonlinearly with the number of cooperators \cite{sheng2024strategy,hauert2006synergy}: 
\begin{equation}
    \begin{aligned}
        a_{n_{C}} &= \frac{b(1 - \delta^{n_{C} + 1})}{\ell(1 - \delta)} - c, \\
        b_{n_{C}} &= \frac{b(1 - \delta^{n_{C}})}{\ell(1 - \delta)},
    \end{aligned}
\end{equation}
where the benefit factor $\delta$ controls the nonlinearity: $\delta = 1$ is the linear case, $\delta > 1$ encodes synergy (each additional cooperator produces more benefit), and $\delta < 1$ encodes antagonism in the production of public benefits.

Before turning to the hypergraph, consider the classical, well-mixed case. In a single one-shot game played among $\ell$ players with no spatial structure, a cooperator's marginal return on its own contribution is $b \delta^{n_{C}}/\ell$ when there are $n_{C}-1$ other cooperators in the group. For the linear public goods game ($\delta = 1$) this marginal return is independent of $n_{C}$, and the game admits a single, sharp Nash threshold $(b/c)^* = \ell$: below it defection is the dominant strategy, above it cooperation is. For $\delta \ne 1$, the well-mixed game has two thresholds rather than one: $(b/c)^* = \ell$, where the all-defector state ceases to be Nash, and $(b/c)^* = \ell/\delta^{\ell - 1}$, where the all-cooperator state becomes Nash. In every case the thresholds grow with group size. The question for the hypergraph is whether spatial structure---and the positive assortment of cooperators it induces---can shift these baselines.

\begin{figure}[t]
\centering
\includegraphics[width=0.9\textwidth]{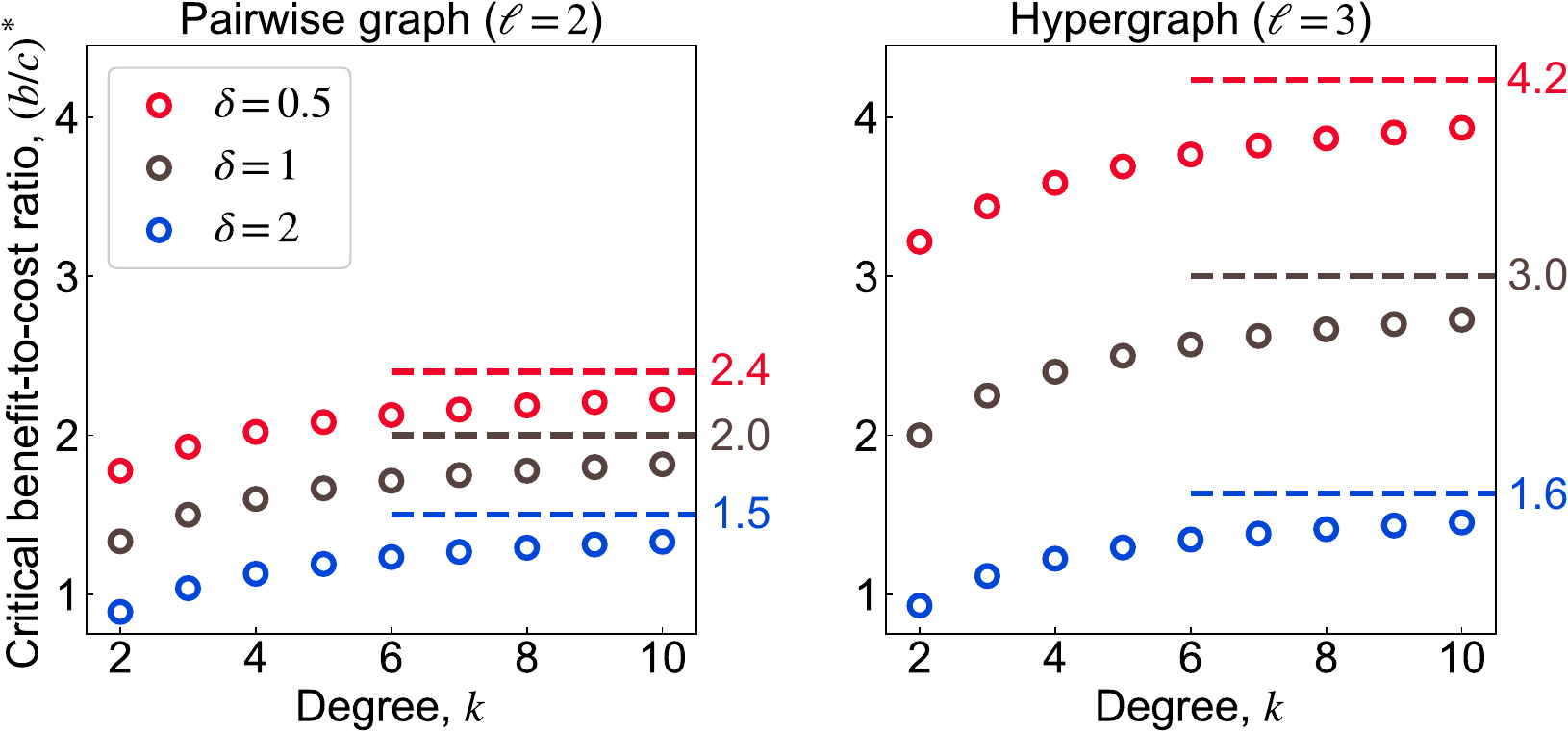}
\caption{\textbf{Threshold for cooperation in public goods games.}
Theoretical predictions for the critical benefit-to-cost ratio, $\left(b/c\right)^{\ast}$, required for cooperation to be favored in the public goods game, for hypergraphs of different orders $\ell$ and degrees $k$. For fixed benefit factor $\delta$, the critical ratio $\left(b/c\right)^{\ast}$ increases monotonically with degree $k$ and approaches a constant as $k \to \infty$, corresponding to the well-mixed limit. For any given degree $k$, the critical ratio is larger for $\ell=3$ than for $\ell=2$, indicating that public goods games played in larger groups generate a stronger social dilemma.
}
\label{fig:pgg}
\end{figure}

On a large regular hypergraph,  substituting the public-goods payoffs into Eqs.~\ref{eq:l2} and \ref{eq:l3} yields a closed-form expression for the critical benefit-to-cost ratio to favor cooperation:
\begin{equation}
    \bcratio = \left\{
    \begin{aligned}
        &\frac{6 k^2}{(k+1) (k(\delta+2)+\delta-1)} \quad &\ell =2, \\[1em]
        &\frac{36 k^2 (2 k-1)}{(k+1) \left(\delta^2 \left(4 k^2+3\right)+\delta \left(8 k^2-6\right)+3 (2 k-1)^2\right)} \quad &\ell =3.
    \end{aligned}
    \right.
\end{equation}
Figure~\ref{fig:pgg} plots the critical ratio $(b/c)^*$ as a function of the degree $k$, for both $\ell = 2$ and $\ell = 3$ and several values of the benefit factor $\delta$.
Note that these critical ratios are derived under the condition for the absolute success ($\rho_C>1/N$), different from those based on the relative success ($\rho_C > \rho_D$) when the game is nonlinear ($\delta \neq 1$).

Two qualitative features stand out.
First, for any fixed $\delta$ and $k$, the threshold for $\ell = 3$ exceeds the threshold for $\ell = 2$: the social dilemma intensifies in larger groups, consistent with the donation-game intuition that a single cooperator's contribution must be shared among more co-players.
Second, in contrast to the donation game, where the threshold $(b/c)^* = k$ diverges with the degree, the public goods threshold remains bounded as $k \to \infty$, converging to $6/(\delta + 2)$ for $\ell = 2$ and to $18/(\delta^2 + 2\delta + 3)$ for $\ell = 3$.
This result reflects the fact that the donation game presents a stronger social dilemma than the public goods game. In the public goods game a cooperator always retains a share of its own contribution, whereas in the donation game a cooperator gains nothing unless another cooperator is present in its hyperedge.

The spatial cooperation threshold has a precise relationship to the well-mixed Nash conditions. For $\ell = 3$ in the well-mixed limit, $\lim_{k \to \infty} (b/c)^* = 18/(\delta^2 + 2\delta + 3)$. This value coincides with the well-mixed Nash boundaries $b/c = \ell$ and $b/c = \ell/\delta^{\ell - 1}$ only at $\delta = 1$, where all three thresholds collapse to $(b/c)^* = \ell = 3$. For $\delta \ne 1$ the spatial threshold lies strictly inside the well-mixed Nash interval and matches neither endpoint. The reason is that the marginal return $b\delta^{n_{C}}/\ell$ depends on the local composition $n_{C}$ whenever $\delta \ne 1$: the well-mixed Nash analysis evaluates this return at extreme compositions ($n_{C} = 0$ for all-defector stability, $n_{C} = \ell - 1$ for all-cooperator stability), whereas the death-birth dynamics on a hypergraph averages it over the quasi-equilibrium distribution of compositions experienced by a focal cooperator versus a focal defector. Spatial structure thus selects a single, sharp cooperation threshold from within the range that the well-mixed Nash analysis can only bracket.

Finite $k$ further lowers the cooperation threshold relative to the well-mixed ($k \to \infty$) limit. For $\delta = 1$, the closed forms collapse to the simple expression $(b/c)^* = \ell k/(k+1)$ for both $\ell = 2$ and $\ell = 3$, which lies strictly below the well-mixed Nash threshold $\ell$. The reduction is largest when $k$ is small: a sparse hypergraph imposes the strongest positive assortment among cooperators, and therefore the greatest discount on the cost of contributing to the public good. The same monotonicity persists for $\delta \ne 1$ (Fig.~\ref{fig:pgg}): for any fixed $\delta$, $(b/c)^*$ decreases as $k$ decreases, so spatial structure always makes cooperation easier than the well-mixed limit alone would predict.

\begin{figure}[t]
\centering
\includegraphics[width=\textwidth]{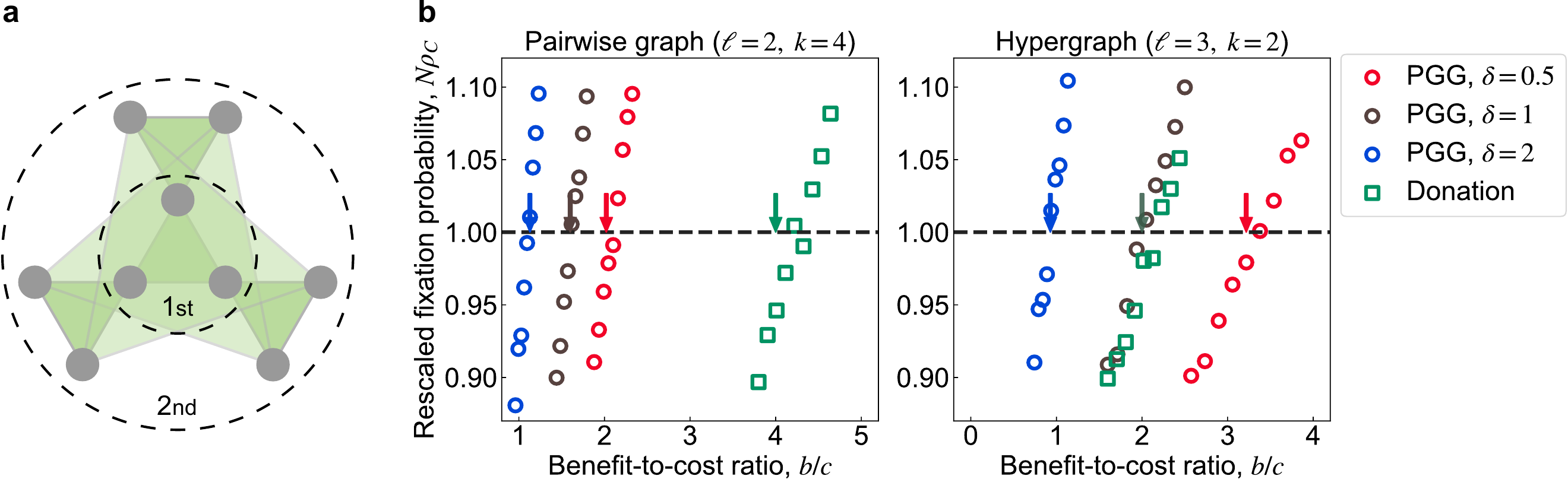}
\caption{\textbf{Monte Carlo simulations confirm theoretical predictions for evolutionary games on hypergraphs.}
\textbf{a}, A vertex-transitive population structure organized into $m$ layers. The $n$th layer contains $3\cdot 2^{n-1}$ nodes, so the total population size is $N = 3(2^m - 1)$. Every pair of nodes in the $n$th layer is connected to a single node in the $(n-1)$th layer. The degree is $k = 4$ for a graph with pairwise interactions ($\ell=2$), and $k = 2$ for a structure with three-way interactions ($\ell=3$).
\textbf{b}, Rescaled fixation probability $N \rho_{C}$ as a function of the benefit-to-cost ratio $b/c$, for the donation game and the public goods game.
Circles denote averages across $2 \times 10^6$ independent Monte Carlo simulations; arrows indicate theoretical predictions from the $\ell$-hyperedge approximation.
Parameters: $m = 5$ ($N = 95$), $c = 1$, $\theta = 0.01$.
}
\label{fig:gamesimu}
\end{figure}

We validate these analytical predictions against Monte Carlo simulations on a vertex-transitive hypergraph (Fig.~\ref{fig:gamesimu}a) in which nodes are organized into $m$ layers, the $n$th layer contains $3 \cdot 2^{n-1}$ nodes, and the total population size is $N = 3(2^m - 1)$.
The  degree is $k = 4$ for $\ell = 2$ and $k = 2$ for $\ell = 3$.
Figure~\ref{fig:gamesimu}b shows that Monte Carlo estimates of $N \rho_{C}$ agree with the $\ell$-hyperedge approximation, for both the donation game and the public goods game, across a range of $b/c$ values.
The approximation also remains accurate on random regular hypergraphs generated by the configuration model and even on heterogeneous hypergraphs whose degree varies across nodes (Supplementary Fig.~2).

\subsection{Complex contagions on hypergraphs}
\subsubsection{Complex contagion without payoff bias}
In the analysis of evolutionary games above, higher-order interactions enter only through payoffs, while state updates still follow a simple contagion process based on pairwise comparisons: that is, the rates in Eq.~\ref{eq:rate_game} are linear in $n_{C}$ and $n_{D}$.
We now turn to complex contagion, where the likelihood that an individual adopts a particular strategy has a nonlinear dependence on how many individuals adopt that strategy within the hyperedge---so that higher-order structure shapes the inheritance process, not just the payoffs.

Following the literature on complex contagion \citep{vasconcelos2019consensus,centola2018behavior}, we first consider the neutral case in which each strategy has the same fitness, so that dynamics arise from the contagion process alone.
In this context, strategies $C$ and $D$ can be interpreted as two competing opinions, but neither opinion has an intrinsic advantage over the other.
Specifically, we posit the following rates for strategic updating,
\begin{equation}
\begin{aligned}
    F_{C}\left(n_{C},n_{D};\theta\right) &= G_{C}\left(n_{C},n_{D};\theta\right) = n_{C}^{1+\lambda \theta}, \\
    F_{D}\left(n_{C},n_{D};\theta\right) &= G_{D}\left(n_{C},n_{D};\theta\right) = n_{D}^{1+\lambda \theta},
\end{aligned}
\label{eq:rates_complex}
\end{equation}
which are controlled by a complexity parameter $\lambda$.
When $\lambda=0$, the dynamics reduce to simple contagion. For $\lambda > 0$ and $\lambda < 0$, the dynamics follow complex contagion, corresponding to conformity-seeking and novelty-seeking behavior, respectively. 

Complex contagion is an inherently group-level phenomenon. The inheritance of states can no longer be represented by parent-to-offspring maps, but rather inheritance arises from the entire composition of states in the group of size $\ell$.
On a pairwise graph, the rates in Eq.~\ref{eq:rates_complex} are either $0$ or $1$ regardless of $\lambda$, so the complexity parameter has no effect on the dynamics; complex contagion requires $\ell \ge 3$.
For the same reason, existing analytical approaches based on parent-to-offspring inheritance and coalescent theory \cite{allen:evolutionary:2017,mcavoy2021fixation,allen2019mathematical,allen2014measures} do not apply to complex processes on hypergraphs. For complex contagion, inheritance is mediated by the hyperedge as a whole rather than by a single parent, and the standard coalescent constructions break down.

\begin{figure}[t]
\centering
\includegraphics[width=\textwidth]{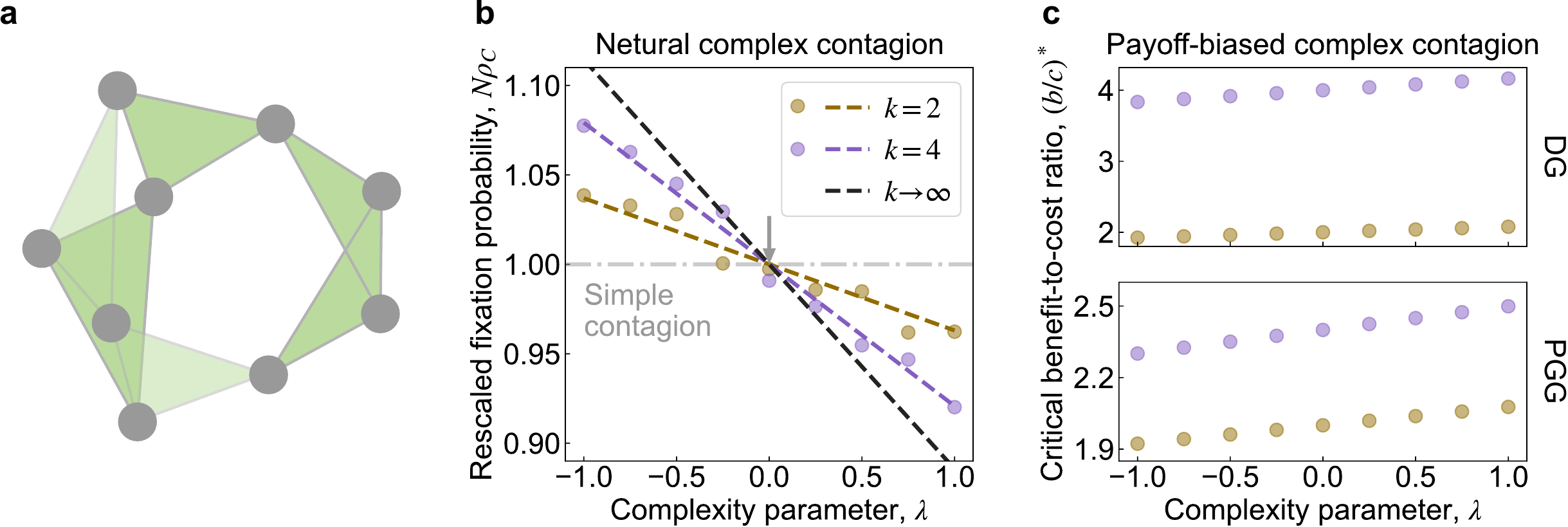}
\caption{\textbf{Complex contagion on hypergraphs.} 
We consider complex contagion dynamics on third-order random regular hypergraphs generated by the configuration model (Algorithm 1 in SI).
\textbf{a}, An illustrative example of a random regular hypergraph of size $N=9$ and degree $k=2$.
\textbf{b}, Rescaled fixation probability, $N\rho_{C}$, under neutral complex contagion, where the two strategies have the same fitness. Dots show results from $2\times 10^6$ independent Monte Carlo simulations, and dashed lines show the theoretical predictions. For fixed degree $k$, the fixation probability decreases monotonically with the complexity parameter $\lambda$: conformity-seeking dynamics $(\lambda>0)$ suppress the spread of a rare type, whereas novelty-seeking dynamics $(\lambda<0)$ promote it. For fixed $\lambda$, the deviation from neutrality becomes stronger as $k$ increases and approaches a limiting value as $k\to\infty$.
\textbf{c}, Effect of complex contagion on cooperation when contagion dynamics are combined with the donation game (DG) or the public goods game (PGG). Increasing $\lambda$ raises the critical ratio $\left(b/c\right)^{\ast}$, indicating that conformity hinders, whereas novelty promotes, the evolution of cooperation.
Parameters: $c=1$, $\theta=0.01$, and $N=99$ for simulations.
}
\label{fig:complexcontagion}
\end{figure}

To quantify the effect of complexity on the spread of a rare type, or opinion, we consider the invasion strength, the leading-order effect of selection on the fixation probability of one type: $\partial_{\theta} \rho_{C} |_{\theta = 0}$. 
In particular, we study how the invasion strength changes with the average degree $k$ and the complexity parameter $\lambda$. 
For $\ell =3$, the invasion strength is
\begin{equation}
    \partial_{\theta}\rho_{C}|_{\theta =0} = -\frac{\lambda \left(\ln 2\right)\left(2k-3\right)}{6\left(2k-1\right)}
    \label{eq:first_order}
\end{equation}
in the large population limit (see Eq.~\ref{eq:prob_complex} in Methods). 
This expression shows that the sign of the invasion strength is determined entirely by $\lambda$: novelty-seeking dynamics $(\lambda<0)$ promote the spread of a rare type, whereas conformity-seeking dynamics $(\lambda>0)$ suppress it. Moreover, for fixed $\lambda$, the magnitude of the invasion strength increases monotonically with degree $k$.
Thus the effect of complex contagion is weakest on sparsely connected hypergraphs and becomes stronger as the population approaches the well-mixed limit. When the average degree approaches that limit, $k\to \infty$, the invasion strength converges to $-\lambda \ln 2/6$. Monte Carlo simulations on random regular hypergraphs (Fig.~\ref{fig:complexcontagion}a,b) confirm both the quantitative predictions in Eq.~\ref{eq:first_order} and these qualitative trends.

These results admit a simple interpretation. When $\lambda>0$, the update rule is conformity-seeking: the effective influence of a strategy grows more than linearly with the number of individuals already holding it in the hyperedge. A rare type therefore faces an additional disadvantage, because it typically appears in groups where it is underrepresented, and so its fixation probability is reduced. When $\lambda<0$, by contrast, the update rule is novelty-seeking: the influence of a strategy grows sublinearly with local abundance, which softens the disadvantage of being rare and increases the chance that a rare type spreads. In this sense, a positive complexity parameter acts like an additional discounting effect on rare opinions, whereas a negative complexity parameter acts like a synergistic effect that favors their initial growth.

\subsubsection{Complex contagion with payoff-biased reproduction}
We now combine the two mechanisms studied above: payoff-biased selection from evolutionary game dynamics and nonlinear state updating from complex contagion. In this setting, higher-order interactions affect both phases of the process. They determine payoffs through the game played within each hyperedge, and they also determine how strongly the local composition of the hyperedge biases subsequent strategy updating. Thus the same group structure shapes both fitness and inheritance.

To model this combined process, we let adoption rates depend multiplicatively on payoff and on the complexity of contagion. The resulting rates are
\begin{equation}
\begin{aligned}
    F_{C}\left(n_{C},n_{D};\theta\right) &= n_{C}^{1+\lambda \theta} \exp\left({\theta u_{C}\left(n_{C},n_{D}\right)}\right),\ F_{D}\left(n_{C},n_{D};\theta\right) = n_{D}^{1+\lambda \theta} \exp\left( \theta u_{D}\left(n_{C},n_{D}\right)\right) ,\\
    G_{C}\left(n_{C},n_{D};\theta\right) &= n_{C}^{1+\lambda \theta} \exp\left(\theta v_{C}\left(n_{C},n_{D}\right)\right),\ G_{D}\left(n_{C},n_{D};\theta\right) = n_{D}^{1+\lambda \theta} \exp\left(\theta v_{D}\left(n_{C},n_{D}\right)\right).
    \label{eq:rate_combine}
\end{aligned}
\end{equation}
Under weak selection, the first-order term of each rate in Eq.~\ref{eq:rate_combine} is simply the sum of the corresponding first-order terms from the pure game-dynamic case (Eq.~\ref{eq:rate_game}) and the pure complex-contagion case (Eq.~\ref{eq:rates_complex}). 

We can understand how game dynamics and complex contagion collectively drive evolution on hypergraphs by analyzing the probability that a focal defector switches to cooperation within a hyperedge $(D,n_C,n_D)$, namely $F_C/(F_C+F_D)$.
A Taylor expansion of this expression in the selection strength $\theta$ produces the following intuitive expression:
\begin{equation}
    \frac{F_C}{F_C + F_D}=\frac{n_C^{1+\lambda \theta} e^{\theta u_C}}
{n_C^{1+\lambda \theta} e^{\theta u_C} + n_D^{1+\lambda \theta} e^{\theta u_D}} \approx \frac{n_C}{\ell - 1} + \theta \, \frac{n_C n_D}{(\ell - 1)^2} \left[(u_C - u_D) + \lambda \ln\left(\frac{n_C}{n_D}\right) \right].
\end{equation}
Here the zeroth-order term $n_C/(\ell-1)$ recovers the classical voter model -- namely, without selection the focal defector adopts cooperation with a probability equal to the fraction of cooperators among the other individuals in the hyper-edge. The strength of selection $\theta$ enters into the first-order term, whose components each have clear interpretations.  The prefactor $n_Cn_D/(\ell-1)^2$ represents the diversity of types in the hyperedge: it is the variance of a Bernoulli random variable with parameter $n_C/(\ell-1)$. This prefactor reflects the fundamental result of Price's equation \cite{price1970selection}: the rate of change of a type's frequency scales with its instantaneous diversity. The term $u_C-u_D$ represents the payoff difference between a cooperator and a defector, which is the source of payoff-driven change; and the term $\lambda\ln(n_C/n_D)$ represents the effect of the payoff-neutral complex contagion, which depends on the skew in strategy frequencies (quantified by $\ln(n_C/n_D)$) modulated by form of complexity, $\lambda$. There is a corresponding, intuitive decomposition of the fixation probability, $\rho_C$, into mechanistically distinct additive components (see Eq.~\ref{eq:prob_comb} in Methods).

We can now ask how complex contagion modifies the threshold for cooperation when higher-order interactions shape both payoff and inheritance. Focusing on \(\ell=3\), we compute the critical benefit-to-cost ratio \(\left(b/c\right)^{\ast}\) for both the donation game and the public goods game as a function of the complexity parameter \(\lambda\). Figure~\ref{fig:complexcontagion}c shows that \(\left(b/c\right)^{\ast}\) increases monotonically with \(\lambda\) (see Eqs.~\ref{eq:bcr_dg_complex} and \ref{eq:bcr_pgg_complex} in Methods). Thus conformity-seeking contagion \((\lambda>0)\) makes cooperation harder to evolve, whereas novelty-seeking contagion \((\lambda<0)\) lowers the threshold and promotes cooperation. This monotonicity follows directly from the fact that increasing \(\lambda\) lowers the fixation probability \(\rho_C\) of a rare cooperator.

We also consider a variant in which the payoffs in Eq.~\ref{eq:rate_combine} are averaged over the \(k\) interactions rather than accumulated (Supplementary Fig.~3). The qualitative effect of complex contagion is unchanged: increasing \(\lambda\) still raises the critical ratio \(\left(b/c\right)^{\ast}\). Quantitatively, however, the dependence on \(\lambda\) is stronger under averaged payoffs, so the shifts in the cooperation threshold are more pronounced than in the accumulated-payoff case.

\section{Discussion}
Pair approximation, and its generalization to $n$-site approximations \cite{ben1992mean}, has long served as the workhorse for modeling dynamical processes on graphs.
But many systems exhibit genuine higher-order interactions that cannot be decomposed into pairs, such as ring-exchange interactions in atomic systems \cite{dai2017four}, social contagion on simplicial structures \cite{iacopini2019simplicial}, and the group processes analyzed here.
The $\ell$-hyperedge approximation we develop closes this gap, extending the analytical reach of pair approximation to stochastic processes on regular hypergraphs of arbitrary order.

We have used the approximation to analyze two classes of dynamics on hypergraphs---evolutionary game dynamics and neutral complex contagion---as well as a coupled model that combines them.
The main results include the generalization of the classical $b/c > k$ cooperation rule to multi-player donation games on hypergraphs of arbitrary order; a closed-form threshold benefit-to-cost ratio for the nonlinear public goods game, which remains bounded as $k \to \infty$; a closed-form expression for the fixation probability under neutral complex contagion that exposes the role of a single complexity parameter; and a characterization of how conformity and innovation jointly reshape the conditions for the evolution of cooperation.

An alternative theoretical framework is evolutionary set theory \cite{tarnita2009evolutionary}, which also describes group interactions in structured populations.
The two approaches differ in how they treat the group.
Evolutionary set theory tracks individuals migrating between groups according to individual fitness; the $\ell$-hyperedge approximation treats the groups themselves as static interaction units, and the dynamics live on the strategies rather than on the group memberships.

Coalescent theory has likewise been applied to evolutionary dynamics on finite structured populations \cite{allen2014measures,su2022multilayer,mcavoy2022evaluating,allen2024coalescent}, and recent advances have extended it to nonlinear cooperation on hypergraphs \cite{sheng2024strategy}.
Coalescent methods are exact, but they require solving a linear system of size $O(N^{\ell+1})$ for each specific hypergraph \cite{sheng2024strategy}, so the computational cost grows dramatically with $\ell$ and becomes infeasible for $\ell \ge 4$.

On that note, it is worth noting that recent mathematical models of behavioral transmission have been captured by replacement rules, which specify a distribution over pairs $\left(R,\alpha\right)$, where $R\subseteq\left\{1,\dots ,N\right\}$ is a set of sites (or individuals) to be replaced and $\alpha :R\rightarrow\left\{1,\dots ,N\right\}$ is the offspring-to-parent map \cite{mcavoy2021fixation}. If $x_{i}\in\left\{C,D\right\}$ represents the type of individual $i$ at a given point in time, then the type of individual $i$ in the next generation is $x_{\alpha\left(i\right)}$ if $i\in R$ (and $x_{i}$ otherwise). Although this model of evolutionary dynamics can represent horizontal transmission, such as imitation dynamics, it is fundamentally predicated on the assumption that an individual inherits its type from some other individual in the population.

However, processes like complex contagion involve joint dependence on multiple individuals in the population and cannot necessarily be represented by an offspring-to-parent map. In the latter, competition for reproduction and survival can be complex but inheritance is simple; in the former, both of these components are complex. To suitably generalize the notion of a replacement process, suppose that $S$ is the space of possible ``types'' (generalizing $\left\{C,D\right\}$) and note that $\alpha :R\rightarrow\left\{1,\dots ,N\right\}$ induces a pullback map $\beta :S^{N}\rightarrow S^{R}$, defined exactly by inheritance. But this is just one way in which such a map arises, so to expand the scope of this kind of general model of evolutionary dynamics, we can substitute pairs $\left(R,\alpha\right)$ with pairs $\left(R,\beta\right)$, where $R\subseteq\left\{1,\dots ,N\right\}$ is the set of updated sites (or individuals) and $\beta :S^{N}\rightarrow S^{R}$ is a map specifying a type for each element of $R$ based on the current population configuration in $S^{N}$. Here, the configuration influences both the probability that $\left(R,\beta\right)$ is chosen and the subsequent behavioral updates.

Although the goal of this paper is not to work out the evolutionary dynamics of general ``adoption rules'' of this form, it is worth noting that the existing analysis of general evolutionary processes based on replacement rules does not contain all adoption rules as special cases. Since this kind of analysis is currently done for replacement rules \cite{ohtsuki:simple:2006,allen:evolutionary:2017,mcavoy2021fixation,sheng2024strategy}, it suffices to show that updates in complex contagion cannot be described by the pullback of offspring-to-parent maps. A simple illustrative example, if pathological, is the case in which there are three types, $A$, $B$, and $C$, and adoption of $C$ is based on joint exposure to $A$ and $B$ in a hyperedge. In the case of $N=3$, with a single $3$-hyperedge in the population, a single-site update of the first individual in $\left(A,A,B\right)$ to $C$ cannot come from a replacement rule, since the only options would be for inheritance from type $A$ or type $B$. In other words, requiring a parent-to-offspring inheritance structure restricts processes to simple contagion and excludes the complex contagion processes we have considered here. Furthermore, $\ell$-hyperedge approximation is non-exact, but it scales to large $\ell$ and accommodates complex contagion as a natural special case.

Our results show that novelty-seeking behavior can promote the spread of a rare strategy or opinion, reducing the synergy required for cooperation to emerge (Supplementary Fig.~4).
In real populations, however, innovation typically incurs additional costs in development and production \cite{dimasi2003price,cuijpers2011costs,chegut2019price}.
A natural extension is to incorporate these costs directly into the payoff structure, yielding a tradeoff between the benefits of novelty and its direct costs, and an optimal bias towards innovation that can itself be characterized within the $\ell$-hyperedge framework.

The examples considered here suggest a broader range of applications. Any stochastic process in which outcomes depend on the composition of a group, rather than on pairwise contacts alone, is a natural candidate for the \(\ell\)-hyperedge approximation. This includes group-mediated transmission processes, as well as evolutionary or cultural dynamics with more than two competing types. In multi-type settings, the approximation can be combined with the standard embedded-Markov-chain approach, using pairwise fixation probabilities to derive the stationary distribution across strategies or opinions.

\section*{Methods}
\subsection*{$\ell$-hyperedge approximation and fixation probability}
The dynamical system can be represented by the variables $p_{C}$, $\qc$, and $\qd$ (for all $n_{C}+n_{D} = \ell-1$).
Since the stochastic process of $p_{C}$ is a martingale under neutral drift ($\theta =0$), the dynamics associated with hyperedges, $\qc$ and $\qd$, unfold on a much faster timescale than that of the individual strategy $p_{C}$ under weak selection ($\theta \ll 1$). As a result, $\qc$ and $\qd$ rapidly reach equilibrium and can be expressed as polynomial functions of $p_{C}$ and the average degree, $k$,
\begin{subequations}
\begin{align}
    \qc &= \left(-1 \right)^{n_{D}} \binom{\ell-1}{n_{D}} \frac{\prod_{i=1}^{n_{C}} \left(i+ h p_{C} \right) \prod_{j=1}^{n_{D}}\left(h p_{C}- \left(h+j-1 \right) \right)}{\prod_{i=1}^{\ell-1} \left(i+h \right)}, \\
    \qd &= \left(-1 \right)^{n_{D}} \binom{\ell-1}{n_{D}} \frac{\prod_{i=1}^{n_{C}} \left(i - 1+ h p_{C} \right) \prod_{j=1}^{n_{D}}\left(h p_{C}- \left(h+j \right) \right)}{\prod_{i=1}^{\ell-1} \left(i+h \right)},
\end{align}
\label{eq:qcqd}
\end{subequations}
where $h=\left(\ell-1 \right)k - \ell$.

A useful way to interpret Eq.~\ref{eq:qcqd} is through an analogy with classical population genetics. The equilibrium expressions for \(q_{(n_C,n_D)\mid C}\) and \(q_{(n_C,n_D)\mid D}\) are not just a convenient closure of the hyperedge dynamics; they belong to the same mathematical family as the stationary mutation--drift distributions that arise in Wright--Fisher theory. In that setting, one thinks of a local allele frequency as fluctuating under genetic drift while being pulled toward an external mean by mutation or, in our case, immigration from a global pool. Here, the analogous quantity is the local frequency of cooperators in a focal hyperedge. The key point is that Eq.~\ref{eq:qcqd} can be read as saying that this local frequency is distributed according to a Beta distribution whose mean is set by the global frequency of cooperators \(p_C\), and whose concentration is set by the single parameter
\[
h=(\ell-1)k-\ell.
\]
Thus \(p_C\) determines the center of the local distribution of cooperators, while \(h\) determines how tightly local composition is concentrated around that center.

To see this connection more explicitly, it is helpful to rewrite Eq.~\ref{eq:qcqd} in rising-factorial form. Since
\begin{equation}
    (-1)^{n_D}\prod_{j=1}^{n_D}\bigl(hp_C-(h+j-1)\bigr)
=
\prod_{j=1}^{n_D}\bigl(h(1-p_C)+j-1\bigr),
\end{equation}
we obtain
\begin{equation}
    q_{(n_C,n_D)\mid C}
=
\binom{\ell-1}{n_D}
\frac{(1+h p_C)_{n_C}\,(h(1-p_C))_{n_D}}{(1+h)_{\ell-1}},
\end{equation}
and similarly
\begin{equation}
    q_{(n_C,n_D)\mid D}
=
\binom{\ell-1}{n_D}
\frac{(h p_C)_{n_C}\,(1+h(1-p_C))_{n_D}}{(1+h)_{\ell-1}},
\end{equation}
where \((x)_m=x(x+1)\cdots(x+m-1)\) denotes the rising factorial. These are exactly Beta-binomial probabilities that often arise in population genetics. Equivalently, if \(X\) denotes a local frequency of cooperators, then
\begin{equation}
    X\mid C \sim \mathrm{Beta}(1+h p_C,\; h(1-p_C)),
\qquad
X\mid D \sim \mathrm{Beta}(h p_C,\; 1+h(1-p_C)),
\end{equation}
and the number \(n_C\) of cooperators among the remaining \(\ell-1\) positions in the hyperedge is obtained by binomial sampling from \(X\). Ignoring for the moment the conditioning on the focal type, the underlying unconditioned form is therefore
\begin{equation}
    X\sim \mathrm{Beta}(h p_C,\; h(1-p_C)),
\end{equation}
which is precisely the standard mean--concentration parameterization of a Beta distribution from population genetics. In this form, the mean is
\[
\mathbb{E}[X]=p_C,
\]
and the variance is
\[
\mathrm{Var}(X)=\frac{p_C(1-p_C)}{h+1}.
\]
Thus \(p_C\) plays the role of the global or mainland frequency, while \(h\) plays the role of the compound immigration--drift parameter that controls the magnitude of local fluctuations: larger \(h\) means weaker genetic drift (relative to immigration) and tighter concentration around the mean local frequency.

The conditional law given a focal cooperator has the same interpretation, but with one additional pseudo-count on the \(C\) side:
\[
X\mid C \sim \mathrm{Beta}(1+h p_C,\; h(1-p_C)).
\]
Thus conditioning on the focal type does not change the basic interpretation of the distribution: \(p_C\) still controls the central tendency of local composition, while \(h\) still controls the strength of concentration around that center. The extra \(1\) in the first Beta parameter simply reflects the fact that, once the focal individual is known to be a cooperator, the local frequency is biased upward by one effective count on the \(C\) side.

In our model, the concentration parameter is
\[
h=(\ell-1)k-\ell,
\]
so both \(k\) and \(\ell\) enter only through this single compound quantity. 
The global frequency \(p_C\) sets the mean local composition in the hyperedge. The parameters \(k\) and \(\ell\) together determine the concentration \(h\) around the mean. Increasing the number of hyper-edges \(k\) increases \(h\), so it strengthens mixing across hyperedges and makes local composition less variable and more tightly tied to the global frequency \(p_C\); this makes intuitive sense because a focal node is involved in more interactions with other players. Increasing the order \(\ell\) also increases \(h\) whenever \(k\ge 2\), so larger interaction groups likewise increase concentration and reduce drift-like fluctuations in local composition. In this sense, Eq.~\ref{eq:qcqd} shows that local hyperedge composition is mathematically identical to a Wright--Fisher mutation--drift equilibrium, reparameterized in terms of the global frequency \(p_C\) and the effective concentration \(h=(\ell-1)k-\ell\). This population-genetic interpretation helps explain why the \(\ell\)-hyperedge approximation takes such a natural and tractable form.

Furthermore, Eq.~\ref{eq:qcqd} can be rewritten as $\qc=\sum_{m=0}^{\ell-1}\alpha^{\left(m\right)}_{n_{C},n_{D}}p_{C}^{m}$ and $\qd=\sum_{m=0}^{\ell-1}\beta^{\left(m\right)}_{n_{C},n_{D}}p_{C}^{m}$.
For $\alpha_{n_{C},n_{D}}^{\left(m\right)}$, we have
\begin{equation}
    \alpha_{n_{C},n_{D}}^{\left(m\right)} = \left(-1 \right)^{n_{D}} \binom{\ell-1}{n_{D}} \frac{h^{m}\left(\prod_{i=1}^{\ell-1} c_i \right) / \left(\sum_{i_1=1}^{\ell-1}\sum_{i_2=i_1+1}^{\ell-1}\cdots\sum_{i_m=i_{m-1}+1}^{\ell-1} \frac{1}{c_{i_1} \cdots c_{i_{m}}} \right)}{\prod_{i=1}^{\ell-1} \left(i+h \right)}.
    \label{eqs:alpha}
\end{equation}
where $\{c_n\}_{n=1}^{\ell-1} = \{1,2,\ldots ,n_{C},-h,\ldots ,-\left(h+n_{D} - 1\right)\}$.
For $\beta^{\left(m\right)}_{n_{C},n_{D}}$, when $n_{C} \ge 1$, we have
\begin{equation}
\begin{aligned}
    \beta_{n_{C},n_{D}}^{(0)} &= 0, \\
    \beta_{n_{C},n_{D}}^{\left(m\right)} &= \left(-1 \right)^{n_{D}} \binom{\ell-1}{n_{D}} \frac{h^{m}\left(\prod_{i=1}^{\ell-2} d_i \right) / \left(\sum_{i_1=1}^{\ell-2}\sum_{i_2=i_1+1}^{\ell-2}\cdots\sum_{i_{m-1}=i_{m-2}+1}^{\ell-2} \frac{1}{d_{i_1} \cdots d_{i_{m-1}}} \right)}{\prod_{i=1}^{\ell-1} \left(i+h \right)}\quad \left(m \ge 1\right),
    \label{eqs:beta1}
\end{aligned}
\end{equation}
where $\{d_n\}_{n=1}^{\ell-2} = \{1,2,\ldots ,n_{C}-1,-\left(h+1 \right),\ldots ,-\left(h+n_{D} \right)\}$. When $n_{C}=0$, we have
\begin{equation}
    \beta_{0,l-1}^{\left(m\right)} = \left(-1 \right)^{m} \frac{h^{m}\left(\prod_{i=1}^{\ell-1} d_i \right) / \left(\sum_{i_1=1}^{\ell-1}\sum_{i_2=i_1+1}^{\ell-1}\cdots\sum_{i_{m}=i_{m-1}+1}^{\ell-1} \frac{1}{d_{i_1} \cdots d_{i_{m}}} \right)}{\prod_{i=1}^{\ell-1} \left(i+h \right)},
    \label{eqs:beta2}
\end{equation}
where $\{d_n\}_{n=1}^{\ell-1} = \{-\left(h+1 \right),\ldots ,-\left(h+n_{D} \right)\}$.

Applying the diffusion approximation and Dynkin's Formula \citep{ohtsuki:simple:2006}, the fixation probability that the population reaches the all-$C$ state from the initial condition $p_{C}(t)|_{t=0} = p_0$ is
\begin{equation}
    \rho(p_0) = p_0 + \theta \left( p_0 \sum_{m=0}^{\ell-1} \frac{1}{(m+1)(m+2)} \overline{\tau}^{\left(m\right)} - \sum_{m=0}^{\ell-1} \frac{1}{(m+1)(m+2) } \overline{\tau}^{\left(m\right)} p_0^{m+2}\right) + O\left(\theta^2 \right).
    \label{eq:generalfixationprob}
\end{equation}
Here, $\overline{\tau}^{\left(m\right)}$ is determined by $\alpha_{n_{C},n_{D}}^{\left(m\right)}$ and $\beta_{n_{C},n_{D}}^{\left(m\right)}$ and represents the $m$th polynomial coefficient of the ratio between the expected change and the variance of the change in the frequency of $C$ per unit time, $2\mathbb{E}(\Delta p_{C})/\text{Var}(\Delta p_{C})$.
We let the initial frequency of $C$ be $p_0=1/N$ and take the limit as $N\to \infty$, obtaining the fixation probability of $C$, $\rho_{C}$, as shown in Eq.~\ref{eq:generalfixationprobability}.

\subsection*{Fixation probability under evolutionary game dynamics}
The expected accumulated payoff of a cooperative neighbor in the hyperedge of type $\left(D,n_{C},n_{D}\right)$ is 
\begin{equation}
    u_{C}\left(n_{C},n_{D}\right) = a_{n_{C}-1} + (k-1)\sum_{n_{C}^{(1)}+n_{D}^{(1)}=\ell-1}  a_{n_{C}^{(1)}} q_{(n_{C}^{(1)},n_{D}^{(1)})\mid C},
\end{equation}
where $a_{n_{C}}$ is the payoff of a focal cooperator obtained from a $\ell$th-order interaction with $n_{C}$ cooperative neighbors.
The expected accumulated payoff of a defector neighbor in this hyperedge is
\begin{equation}
    u_{D}\left(n_{C},n_{D}\right) = b_{n_{C}} + (k-1)\sum_{n_{C}^{(1)}+n_{D}^{(1)}=\ell-1}  b_{n_{C}^{(1)}} q_{(n_{C}^{(1)},n_{D}^{(1)})\mid D},
\end{equation}
where $b_{n_{C}}$ is the corresponding payoff of a focal defector.
Similarly, the expected accumulated payoff of a cooperative neighbor in the hyperedge of type $\left(C,n_{C},n_{D}\right)$ is 
\begin{equation}
    v_{C}\left(n_{C},n_{D}\right) = a_{n_{C}} + (k-1)\sum_{n_{C}^{(1)}+n_{D}^{(1)}=\ell-1}  a_{n_{C}^{(1)}} q_{(n_{C}^{(1)},n_{D}^{(1)})\mid C},
\end{equation}
and the expected accumulated payoff of a defector neighbor in this hyperedge is
\begin{equation}
    v_{D}\left(n_{C},n_{D}\right) = b_{n_{C}+1} + (k-1)\sum_{n_{C}^{(1)}+n_{D}^{(1)}=\ell-1}  b_{n_{C}^{(1)}} q_{(n_{C}^{(1)},n_{D}^{(1)})\mid D}.
\end{equation}
In this case, the quantity $\overline{\tau}^{\left(m\right)}$ in Eq.~\ref{eq:generalfixationprob} becomes $\overline{\tau}^{\left(m\right)} = \overline{\mu}^{\left(m\right)} - \overline{v}^{\left(m\right)}$, where
\begin{equation}
    \begin{aligned}
    \gamma_{n_{C},n_{D}}^{\left(m\right)} &\coloneqq \sum_{i=0}^m \alpha_{n_{C},n_{D}}^{\left(i\right)} \quad (0\le m \le \ell-1), \\
        \mu^{\left(m\right)} &\coloneqq \sum_{n_{C}+n_{D} = \ell-1, n_{D} \ge 1} n_{D} \left(a_{n_{C}} -b_{n_{C}+1} \right) \gamma_{n_{C},n_{D}}^{\left(m\right)}\quad (0\le m \le \ell-2),\ \mu^{(\ell-1)} \coloneqq 0, \\
        v^{\left(m\right)} &\coloneqq \summ b_{n_{C}} \beta_{n_{C},n_{D}}^{\left(m\right)} - a_{n_{C}} \alpha_{n_{C},n_{D}}^{\left(m\right)} \quad (0\le m \le \ell-1), \\
        \overline{\mu}^{\left(m\right)} &\coloneqq \frac{N(k-1)}{k\left((\ell-1)k-\ell\right)} \mu^{\left(m\right)}, \\
        \overline{v}^{\left(m\right)} &\coloneqq \frac{N(k+1)(k-1)}{k} v^{\left(m\right)}.
        \label{eq:quantity_game}
    \end{aligned}
\end{equation}
The quantity $\overline{\mu}^{\left(m\right)}$ corresponds to the difference in the expected accumulated payoff between all cooperators surrounded by a defector and all defectors surrounded by a cooperator, while $\overline{v}^{\left(m\right)}$ corresponds to the difference in the expected payoff between a single cooperator and a single defector.
When the population size is large, the fixation probability of cooperation in this case becomes
\begin{equation}
    \rho_{C} = \frac{1}{N} + \frac{\theta}{N} \sum_{m=0}^{\ell-1} \frac{1}{(m+1)(m+2)}\left( \overline{\mu}^{\left(m\right)} - \overline{v}^{\left(m\right)}\right) + O\left(\theta^2\right),
    \label{eq:prob_game}
\end{equation}
and $\rho_{C} > 1/N$ is equivalent to
\begin{equation}
\sum_{m=0}^{\ell-1} \frac{1}{(m+1)(m+2)}\left( \overline{\mu}^{\left(m\right)} - \overline{v}^{\left(m\right)}\right) > 0. 
\label{eq:condition_game}
\end{equation}

\subsection*{$\ell$-player donation game}
When $a_{n_{C}}=-c + n_{C} b$ and $b_{n_{C}}=n_{C} b$, we can prove that 
\begin{equation}
    \frac{2 \mathbb{E}(\Delta p_{C})}{\text{Var}(\Delta p_{C})} = \theta N \left(b- kc \right),
    \label{eq:26}
\end{equation}
which means $\overline{\tau}^{(0)} = \overline{\mu}^{(0)} - \overline{v}^{(0)} = N\left(b-kc \right)$ and $\overline{\tau}^{\left(m\right)} = \overline{\mu}^{\left(m\right)} - \overline{v}^{\left(m\right)} = 0$ for $1\le m \le \ell-1$.
Substituting $\overline{\tau}^{\left(m\right)}$ into Eq.~\ref{eq:condition_game} gives the $b/c>k$ rule on hypergraphs of order $\ell$.

\subsection*{Relative success of cooperation}
The relative success of cooperation over defection is quantified by whether the fixation probability of cooperation exceeds that of defection, i.e., $\rho_{C} > \rho_{D}$. 
The fixation probability of $D$, $\rho_{D}$, can be obtained using the complementary relation $\rho_{D}=1-\rho(1-1/N)$.
Then $\rho_{C} > \rho_{D}$ is equivalent to 
\begin{equation}
\sum_{m=0}^{\ell-1} \frac{1}{m+1}\left( \overline{\mu}^{\left(m\right)} - \overline{v}^{\left(m\right)}\right) > 0,
\label{eq:relativesuccess}
\end{equation}
in the large population limit. 

Take $\ell =2$ and $\ell =3$ as examples.
In the case of $\ell =2$ \cite{ohtsuki:simple:2006}, the condition for $\rho_{C} > \rho_{D}$ is 
\begin{equation}
    \left(k+1 \right) a_0 + \left(k-1 \right)a_1 > (k+1)b_0 + (k-1)b_1. 
\end{equation}
In the case of $\ell =3$, the condition becomes
\begin{equation}
\begin{aligned}
    &\left(8 k^3-10 k^2+3 k+3\right)a_0+ \left(8 k^3-4 k^2-6\right)a_1+ \left(8 k^3+2 k^2-3 k+3\right)a_2 \\
    &>  \left(8 k^3+2 k^2-3 k+3\right)b_0+ \left(8 k^3-4 k^2-6\right)b_1+ \left(8 k^3-10 k^2+3 k+3\right)b_2.
\end{aligned}
\end{equation}

\subsection*{Neutral complex contagion}
For $\ell =3$, the fixation probability of $C$ is
\begin{equation}
    \rho_{C} = \frac{1}{N}+\theta\lambda\left(\ln 2\right) \frac{2k-3}{2k-1} \left(-\frac{1}{6} + \frac{1}{2N} - \frac{1}{3N^2} \right) + O\left(\theta^2 \right).
        \label{eq:prob_complex}
\end{equation}
In the limit $N\to\infty$, $\rho_{C}$ becomes
\begin{equation}
    \rho_{C} = \frac{1}{N}-\frac{\theta \lambda \left(\ln 2\right) (2k-3)  }{6(2 k-1)} + O\left(\theta^2 \right).
\end{equation}
We refer to SI for the analytical derivations of $\rho_{C}$ for arbitrary orders $\ell$. 

\subsection*{Complex contagion with payoff-biased selection}
We calculate the fixation probability of cooperation in the case of evolutionary game dynamics combined with complex contagion for $\ell =3$:
\begin{equation}
\begin{aligned}
        \rho_{C} = \frac{1}{N} + \frac{\theta}{24k(2k-1)}&\left(\substack{a_0 \left(12 k^3-16 k^2+3 k+3\right)+a_1 \left(8 k^3-6 k-6\right)+a_2 \left(4 k^3+4 k^2+3 k+3\right)\\ -b_0 \left(12 k^3-9 k+3\right)-b_1 \left(8 k^3-8 k^2+6 k-6\right)-b_2 \left(4 k^3-4 k^2+3 k+3\right)} \right. \\
        &\left. -4\lambda k\left(\ln 2\right)\left(2k-3\right) \right) + O\left(\theta^2 \right).
\end{aligned}
\label{eq:prob_comb}
\end{equation}

When $c=1$, the critical benefit-to-cost ratio in the donation game is 
\begin{equation}
    \bcratio = k+\frac{\lambda\left(\ln 2\right)\left(2k-3\right)}{3\left(2k-1\right)},
    \label{eq:bcr_dg_complex}
\end{equation}
and the critical benefit-to-cost ratio in the public goods game with benefit factor $\delta_3$ is
\begin{equation}
    \bcratio = \frac{36 k^2 (2 k-1) + 12\lambda k\left(\ln 2\right)\left(2k-3\right)}{\left(k+1\right)\left(\delta_3^2 \left(4 k^2+3\right)+\delta_3 \left(8 k^2-6\right)+3 (2 k-1)^2\right)}.
    \label{eq:bcr_pgg_complex}
\end{equation}

\section*{Acknowledgments}
We thank Hisashi Ohtsuki for insightful comments and suggestions.

\bibliographystyle{unsrtnat}



\section*{Supplementary material}

# Hyperedge approximation for stochastic processes on higher-order networks

*Supplementary Information*

## 1 $\ell$-hyperedge approximation

We consider a population of size $N$. The population structure is represented by a regular graph of order exactly $\ell$ and degree $k$, which means that every individual has exactly $k$ (hyper)edges, each of order $\ell$. Note that the (hyper)graph has a fixed order, so that it does not simultaneously contain, eg, edges and 2-simplices. Every individual chooses either strategy $C$ or $D$ and interacts with their neighbors.

We use the $\ell$-hyperedge approximation, a generalization of the classical pair approximation (1), to analyze stochastic processes on hypergraphs of any given order $\ell$. Let $p_C$ (resp. $p_D$) denote the frequency of $C$-individuals (resp. $D$-individuals) in the population. Let $p_{C,n_C,n_D}$ (resp. $p_{D,n_C,n_D}$) denote the frequency of the hyperedges where the focal individual is a $C$-individual (resp. a $D$-individual) and has $n_C$ $C$-neighbors and $n_D$ $D$-neighbors. We consistently have $n_C + n_D = \ell - 1$. Let $q_{(n_C,n_D)|C}$ (resp. $q_{(n_C,n_D)|D}$) denote the conditional probability that, given the focal individual chooses strategy $C$ (resp. strategy $D$), there are $n_C$ $C$-neighbors and $n_D$ $D$-neighbors in the hyperedge. We have the following identities:

$$p_C + p_D = 1, \tag{1a}$$

$$p_{C,n_C,n_D} = p_C \cdot q_{(n_C,n_D)|C}, \; p_{D,n_C,n_D} = p_D \cdot q_{(n_C,n_D)|D}, \tag{1b}$$

$$\sum_{n_C+n_D=\ell-1} q_{(n_C,n_D)|C} = 1, \quad \sum_{n_C+n_D=\ell-1} q_{(n_C,n_D)|D} = 1, \tag{1c}$$

$$n_D p_{C,n_C,n_D} = (n_C+1) p_{D,n_C+1,n_D-1} \quad (n_D \geq 1)\,. \tag{1d}$$

Equation 1d follows from the fact that a hyperedge containing $n_C + 1$ $C$-individuals and $n_D$ $D$-individuals indicates $n_C + 1$ hyperedges of type $(C, n_C, n_D)$ and $n_D$ hyperedges of type $(D, n_C + 1, n_D - 1)$. By Eq. 1d, we have

$$p_C n_D q_{(n_C,n_D)|C} = p_D \cdot (n_C+1)\, q_{(n_C+1,n_D-1)|D}. \tag{2}$$

Summing both sides of Eq. 2 over $n_D$ from 1 to $\ell - 1$ gives

$$p_C \sum_{n_C+n_D=\ell-1} n_D q_{(n_C,n_D)|C} = p_D \sum_{n_C+n_D=\ell-1} n_C q_{(n_C,n_D)|D}. \tag{3}$$

Note that we introduce $4\ell + 2$ variables and obtain $3\ell + 2$ identities, so the system can be described by $\ell$ independent variables, $p_C$ and $q_{(n_C,n_D)|C}$ ($n_C = 1, ..., \ell - 1$).

### 1.1 Updating a $D$-individual

At each time step, a randomly selected individual has a chance to update their strategy. We first analyze the case where a randomly selected $D$-individual updates their strategy to $C$. Let $k_{n_C,n_D}$ denote the number of hyperedges where there exist $n_C$ $C$-neighbors and $n_D$ $D$-neighbors. We have $\sum_{n_C+n_D=\ell-1} k_{n_C,n_D} = k$. Let $\mathbf{k} = (k_{0,\ell-1}, ..., k_{\ell-1,0})$ denote the neighborhood configuration. The probability that the configuration $\mathbf{k}$ occurs follows a multinomial distribution,

$$\mathcal{M}_D(\mathbf{k}) = \frac{k!}{\prod_{n_C+n_D=\ell-1} k_{n_C,n_D}!} \prod_{n_C+n_D=\ell-1} q_{(n_C,n_D)|D}^{k_{n_C,n_D}}. \tag{4}$$

Let $F_C(n_C, n_D; \theta)$ (resp. $F_D(n_C, n_D; \theta)$) denote the rate that a focal $D$-individual adopts strategy $C$ (resp. strategy $D$) within the hyperedge $(D, n_C, n_D)$ during the updating process. Here, the parameter $\theta$ quantifies the intensity of selection. The probability that the focal $D$-individual switches their strategy to $C$ under configuration $\mathbf{k}$ is obtained

by averaging over all hyperedges:

$$\mathcal{I}_D(\mathbf{k}) = \frac{\sum_{n_C+n_D=\ell-1} k_{n_C,n_D} F_C(n_C, n_D; \theta)}{\sum_{n_C+n_D=\ell-1} k_{n_C,n_D} \left(F_C(n_C, n_D; \theta) + F_D(n_C, n_D; \theta)\right)}, \tag{5}$$

Then, $p_C$ increases by $1/N$ with probability

$$\mathbb{P}\left(\Delta p_C = \frac{1}{N}\right) = p_D \sum_{k_{0,\ell-1}+\ldots+k_{\ell-1,0}=k} \mathcal{M}_D(\mathbf{k})\mathcal{I}_D(\mathbf{k}). \tag{6}$$

Accordingly, $p_{C,(n_C,n_D)}$ increases by $k_{n_C,n_D}/\left(Nk/\ell\right)$ with probability

$$\mathbb{P}\left(\Delta p_{C,(n_C,n_D)} = \frac{\ell k_{n_C,n_D}}{Nk}\right) = p_D \mathcal{M}_D(\mathbf{k})\mathcal{I}_D(\mathbf{k}). \tag{7}$$

## 1.2 Updating a $C$-individual

We then analyze the case where a randomly selected $C$-individual updates their strategy to $D$. The probability that the configuration $\mathbf{k}$ occurs in this case is

$$\mathcal{M}_C(\mathbf{k}) = \frac{k!}{\prod_{n_C+n_D=\ell-1} k_{n_C,n_D}!} \prod_{n_C+n_D=\ell-1} q_{(n_C,n_D)|C}^{k_{n_C,n_D}}. \tag{8}$$

Let $G_C(n_C, n_D; \theta)$ (resp. $G_D(n_C, n_D; \theta)$) denote the rate that a focal $C$-individual adopts strategy $C$ (resp. strategy $D$) within the hyperedge $(C, n_C, n_D)$ during the updating process. The probability that the focal $C$-individual switches their strategy to $D$ under configuration $\mathbf{k}$ is

$$\mathcal{I}_C(\mathbf{k}) = \frac{\sum_{n_C+n_D=\ell-1} k_{n_C,n_D} G_D(n_C, n_D; \theta)}{\sum_{n_C+n_D=\ell-1} k_{n_C,n_D} \left(G_C(n_C, n_D; \theta) + G_D(n_C, n_D; \theta)\right)}. \tag{9}$$

Then, $p_C$ decreases by $1/N$ with probability

$$\mathbb{P}\left(\Delta p_C = -\frac{1}{N}\right) = p_C \sum_{k_{0,\ell-1}+\ldots+k_{\ell-1,0}=k} \mathcal{M}_C(\mathbf{k})\mathcal{I}_C(\mathbf{k}). \tag{10}$$

Accordingly, $p_{C,(n_C,n_D)}$ decreases by $k_{n_C,n_D}/\left(Nk/\ell\right)$ with probability

$$\mathbb{P}\left(\Delta p_{C,(n_C,n_D)} = -\frac{\ell k_{n_C,n_D}}{Nk}\right) = p_C \mathcal{M}_C(\mathbf{k})\mathcal{I}_C(\mathbf{k}). \tag{11}$$

We assume that the evolutionary dynamics of strategies follow a martingale without selection ($\theta = 0$), indicating that the expected change in frequency is unbiased for both strategies $C$ and $D$ at every time step without selection (given the previous strategy frequency $p_C$). Then, we have the identity

$$\frac{F_C(n_C, n_D; 0)}{F_D(n_C, n_D; 0)} = \frac{G_C(n_C, n_D; 0)}{G_D(n_C, n_D; 0)} = \frac{n_C}{n_D}. \tag{12}$$

### 1.3 Differential equations of strategies on nodes and along hyperedges

As only one replacement event occurs in one unit of time, we have the time derivative of $p_C$

$$\dot{p}_C = \frac{1}{N} \cdot \mathbb{P}\left(\Delta p_C = \frac{1}{N}\right) + \frac{1}{N} \cdot \mathbb{P}\left(\Delta p_C = -\frac{1}{N}\right). \tag{13}$$

By Eqs. 3 and 12, we obtain that $\dot{p}_C = 0$ without selection (i.e. $\theta = 0$). Therefore, under weak selection (i.e. $\theta \ll 1$), we have

$$\begin{aligned}\dot{p}_C = \frac{\theta}{N} \Bigg\{ & \frac{\begin{pmatrix} p_D \sum_{n_C+n_D=\ell-1} q_{(n_C,n_D)|D} \partial_\theta F_C(n_C,n_D;0) \\ -p_C \sum_{n_C+n_D=\ell-1} q_{(n_C,n_D)|C} \partial_\theta G_D(n_C,n_D;0) \end{pmatrix}}{\ell-1} - \frac{1}{k(\ell-1)^2} \Bigg[ \begin{pmatrix} p_D \sum_{n_C+n_D=\ell-1} q_{(n_C,n_D)|D} n_C (\partial_\theta F_C(n_C,n_D;0) + \partial_\theta F_D(n_C,n_D;0)) \\ -p_C \sum_{n_C+n_D=\ell-1} q_{(n_C,n_D)|C} n_D (\partial_\theta G_C(n_C,n_D;0) + \partial_\theta G_D(n_C,n_D;0)) \end{pmatrix} \\ & + \begin{pmatrix} (k-1) p_D \sum_{n_C^{(1)}+n_D^{(1)}=\ell-1} n_C^{(1)} q_{n_C^{(1)},n_D^{(1)}|D} \\ \times \sum_{n_C^{(2)}+n_D^{(2)}=\ell-1} q_{n_C^{(2)},n_D^{(2)}|D} \left(\partial_\theta F_C\left(n_C^{(2)},n_D^{(2)};0\right) + \partial_\theta F_D\left(n_C^{(2)},n_D^{(2)};0\right)\right) \\ -(k-1) p_C \sum_{n_C^{(1)}+n_D^{(1)}=\ell-1} n_D^{(1)} q_{n_C^{(1)},n_D^{(1)}|C} \\ \times \sum_{n_C^{(2)}+n_D^{(2)}=\ell-1} q_{n_C^{(2)},n_D^{(2)}|C} \left(\partial_\theta G_C\left(n_C^{(2)},n_D^{(2)};0\right) + \partial_\theta G_D\left(n_C^{(2)},n_D^{(2)};0\right)\right) \end{pmatrix} \Bigg] \Bigg\} + O\left(\theta^2\right).\end{aligned} \tag{14}$$

To simplify this equation, we first prove the following two identities:

$$\begin{aligned} p_D \sum_{n_C+n_D=\ell-1} q_{(n_C,n_D)|D} \partial_\theta F_C(n_C, n_D; 0) &= p_C \sum_{n_C+n_D=\ell-1} \frac{n_D}{n_C+1} q_{(n_C,n_D)|C} \partial_\theta F_C(n_C+1, n_D-1; 0), \\ p_D \sum_{n_C+n_D=\ell-1} q_{(n_C,n_D)|D} n_C \partial_\theta F_C(n_C, n_D; 0) &= p_C \sum_{n_C+n_D=\ell-1} q_{(n_C,n_D)|C} n_D \partial_\theta F_C(n_C+1, n_D-1; 0). \end{aligned} \tag{15}$$

For the first identity, by Eq. 2, we have

$$\begin{aligned} p_D \sum_{n_C+n_D=\ell-1} q_{(n_C,n_D)|D} \partial_\theta F_C(n_C, n_D; 0) &= p_D \sum_{n_C=1}^{\ell-1} q_{(n_C,n_D)|D} n_C \frac{\partial_\theta F_C(n_C, n_D; 0)}{n_C} \\ &= p_D \sum_{n_C=0}^{\ell-2} q_{(n_C+1,n_D-1)|D} (n_C+1) \frac{\partial_\theta F_C(n_C+1, n_D-1; 0)}{n_C+1} \\ &= p_C \sum_{n_C=0}^{\ell-2} q_{(n_C,n_D)|C} n_D \frac{\partial_\theta F_C(n_C+1, n_D-1; 0)}{n_C+1} \\ &= p_C \sum_{n_C+n_D=\ell-1} n_D q_{(n_C,n_D)|C} \frac{\partial_\theta F_C(n_C+1, n_D-1; 0)}{n_C+1} \end{aligned}. \tag{16}$$

We can use a similar idea to prove the second identity.

Define

$$z_C := \sum_{n_C+n_D=\ell-1} n_D q_{(n_C,n_D)|C}, \; z_D = \sum_{n_C+n_D=\ell-1} n_C q_{(n_C,n_D)|D}. \tag{17}$$

Using Eqs. 15 and 17, the time derivative of $p_C$ (Eq. 18) can be simplified as

$$\dot{p}_C = \frac{\theta}{N} \left\{ \frac{p_C \sum_{n_C+n_D=\ell-1} q_{(n_C,n_D)|C} \left( \partial_\theta F_C (n_C+1, n_D-1; 0) \frac{n_D}{n_C+1} - \partial_\theta G_D (n_C, n_D; 0) \right)}{\ell - 1} \right.$$
$$- \frac{1}{k(\ell-1)^2} \left[ \left( p_C \sum_{n_C+n_D=\ell-1} q_{(n_C,n_D)|C} n_D \begin{pmatrix} \partial_\theta F_C(n_C+1,n_D-1;0)+\partial_\theta F_D(n_C+1,n_D-1;0) \\ -\partial_\theta G_C(n_C,n_D;0)-\partial_\theta G_D(n_C,n_D;0) \end{pmatrix} \right) \right. \tag{18}$$
$$\left. \left. + \begin{pmatrix} (k-1)p_D z_D \sum_{n_C+n_D=\ell-1} q_{(n_C,n_D)|D} (\partial_\theta F_C(n_C,n_D;0)+\partial_\theta F_D(n_C,n_D;0)) \\ -(k-1)p_C z_C \sum_{n_C+n_D=\ell-1} q_{(n_C,n_D)|C} (\partial_\theta G_C(n_C,n_D;0)+\partial_\theta G_D(n_C,n_D;0)) \end{pmatrix} \right] \right\} + O\left(\theta^2\right).$$

The time derivative of $p_{C,(n_C,n_D)}$ (for $n_C = 0, ..., \ell - 1$) is

$$\begin{aligned} \dot{p}_{C,(n_C,n_D)} &= p_D \sum_{\mathbf{k}} \frac{\ell k_{n_C,n_D}}{Nk} \mathcal{M}_D(\mathbf{k}) \frac{\sum_{n_C^{(1)}+n_D^{(1)}=\ell-1} k_{n_C^{(1)},n_D^{(1)}} n_C^{(1)}}{k(\ell-1)} \\ &\quad - p_C \sum_{\mathbf{k}} \frac{\ell k_{n_C,n_D}}{Nk} \mathcal{M}_C(\mathbf{k}) \frac{\sum_{n_C^{(1)}+n_D^{(1)}=\ell-1} k_{n_C^{(1)},n_D^{(1)}} n_D^{(1)}}{k(\ell-1)} + O(\theta) \\ &= \frac{\ell p_D}{Nk(\ell-1)} q_{(n_C,n_D)|D} \left( n_C + (k-1) \sum_{n_C^{(1)}+n_D^{(1)}=\ell-1} n_C^{(1)} q_{n_C^{(1)},n_D^{(1)}|D} \right) \\ &\quad - \frac{\ell p_C}{Nk(\ell-1)} q_{(n_C,n_D)|C} \left( n_D + (k-1) \sum_{n_C^{(1)}+n_D^{(1)}=\ell-1} n_D^{(1)} q_{n_C^{(1)},n_D^{(1)}|C} \right) + O(\theta) \end{aligned} . \tag{19}$$

Under weak selection, the frequencies of hyperedges, $p_{C,n_C,n_D}$, reach the equilibrium much faster than the global strategy frequency, $p_C$, that is, $p_{C,n_C,n_D}$ can be expressed as a function of $p_C$. Define

$$x_i := q_{\ell-i,i-1|C} - q_{\ell-i,i-1|D}. \tag{20}$$

By Eq. 1, we have $\sum_{i=1}^{\ell} x_i = 0$ and $p_C z_C = p_D z_D$. Using Eqs. 2 and 3, the equilibrium can be obtained by solving the following system of equations:

$$\begin{cases} p_C \cdot 1 \cdot q_{\ell-2,1|C} - p_C \cdot 0 \cdot q_{\ell-1,0|C} = p_C(k-1)x_1 z_C, \\ p_C \cdot 2 \cdot q_{\ell-3,2|D} - p_C \cdot 1 \cdot q_{\ell-2,1|C} = p_C(k-1)x_2 z_C, \\ \vdots \\ p_C \cdot (\ell-1) \cdot q_{0,\ell-1|C} - p_C \cdot (\ell-2) \cdot q_{1,\ell-2|C} = p_C(k-1)x_{\ell-1} z_C, \\ x_1 \cdot (\ell-1) \cdot (1-p_C) = (\ell-1) \cdot (1-p_C) \cdot q_{\ell-1,0|C} - 1 \cdot p_C \cdot q_{\ell-2,1|C}, \\ x_2 \cdot (\ell-2) \cdot (1-p_C) = (\ell-2) \cdot (1-p_C) \cdot q_{\ell-2,1|C} - 2 \cdot p_C \cdot q_{\ell-3,2|C}, \\ \vdots \\ x_{\ell-1} \cdot 1 \cdot (1-p_C) = 1 \cdot (1-p_C) \cdot q_{1,\ell-2|C} - (\ell-1) \cdot p_C \cdot q_{0,\ell-1|C}, \\ p_C n_D q_{(n_C,n_D)|C} = p_D (n_C+1) q_{(n_C,n_D)|D} \quad n_D = 1, ..., \ell-1, \\ \sum_{n_C+n_D=\ell-1} q_{(n_C,n_D)|C} = 1. \end{cases} \tag{21}$$

As a result, $q_{(n_C,n_D)|C}$ and $q_{(n_C,n_D)|D}$ for all $n_C, n_D$ with $n_C + n_D = \ell - 1$ can be expressed as functions of $p_C$ and $k$.

Let $h = (\ell - 1) k - \ell$. We derive the closed-form solution for the linear system in Eq. 21

$$q_{(n_C,n_D)|C} = (-1)^{n_D} \binom{\ell-1}{n_D} \frac{\prod_{i=1}^{n_C} (i + h p_C) \prod_{j=1}^{n_D} (h p_C - (h + j - 1))}{\prod_{i=1}^{\ell-1} (i + h)}, \tag{22a}$$

$$q_{(n_C,n_D)|D} = (-1)^{n_D} \binom{\ell-1}{n_D} \frac{\prod_{i=1}^{n_C} (i - 1 + h p_C) \prod_{j=1}^{n_D} (h p_C - (h + j))}{\prod_{i=1}^{\ell-1} (i + h)}. \tag{22b}$$

We note that $q_{(n_C,n_D)|C}$ and $q_{(n_C,n_D)|D}$ are polynomial functions of $p_C$ with degree at most $\ell - 1$ for all $n_C, n_D$. It follows that $q_{(n_C,n_D)|C}$ and $q_{(n_C,n_D)|D}$ take the following polynomial forms:

$$q_{(n_C,n_D)|C} = \sum_{m=0}^{\ell-1} \alpha_{n_C,n_D}^{(m)} p_C^m, \; q_{(n_C,n_D)|D} = \sum_{m=0}^{\ell-1} \beta_{n_C,n_D}^{(m)} p_C^m. \tag{23}$$

Let $\{c_n\}_{n=1}^{\ell-1} = \{1, 2, ..., n_C, -h, ..., -(h + n_D - 1)\}$. Comparing Eq. 23 with Eq. 22a gives

$$\alpha_{n_C,n_D}^{(m)} = (-1)^{n_D} \binom{\ell-1}{n_D} \frac{h^m \left(\prod_{i=1}^{\ell-1} c_i\right) / \left(\sum_{i_1=1}^{\ell-1} \sum_{i_2=i_1+1}^{\ell-1} \cdots \sum_{i_m=i_{m-1}+1}^{\ell-1} \frac{1}{c_{i_1} \ldots c_{i_m}}\right)}{\prod_{i=1}^{\ell-1} (i + h)}. \tag{24}$$

For $\beta_{n_C,n_D}^{(m)}$, if $n_C \geq 1$ (i.e. $n_D \leq \ell - 2$), define a sequence $\{d_n\}_{n=1}^{\ell-2} = \{1, 2, ..., n_C - 1, -(h+1), ..., -(h + n_D)\}$. In this case,

$$\begin{aligned} \beta_{n_C,n_D}^{(0)} &= 0, \\ \beta_{n_C,n_D}^{(m)} &= (-1)^{n_D} \binom{\ell-1}{n_D} \frac{h^m \left(\prod_{i=1}^{\ell-2} d_i\right) / \left(\sum_{i_1=1}^{\ell-2} \sum_{i_2=i_1+1}^{\ell-2} \cdots \sum_{i_{m-1}=i_{m-2}+1}^{\ell-2} \frac{1}{d_{i_1} \ldots d_{i_{m-1}}}\right)}{\prod_{i=1}^{\ell-1} (i + h)} \quad (m \geq 1). \end{aligned} \tag{25}$$

If $n_C = 0$ (i.e. $n_D = \ell - 1$), define a sequence $\{d_n\}_{n=1}^{\ell-1} = \{-(h+1), ..., -(h + n_D)\}$. Then, we have

$$\beta_{0,l-1}^{(m)} = (-1)^m \frac{h^m \left(\prod_{i=1}^{\ell-1} d_i\right) / \left(\sum_{i_1=1}^{\ell-1} \sum_{i_2=i_1+1}^{\ell-1} \cdots \sum_{i_m=i_{m-1}+1}^{\ell-1} \frac{1}{d_{i_1} \ldots d_{i_m}}\right)}{\prod_{i=1}^{\ell-1} (i + h)}. \tag{26}$$

Plugging $k$ back in, we obtain the results expressed as functions of $p_C$ and $k$.

Plugging Eq. 22 into Eq. 18 gives

$$\mathbb{E}(\Delta p_C) = \dot{p}_C = \theta H(p_C; k) + O\left(\theta^2\right), \tag{27}$$

where $H$ is a polynomial function of $p_C$.

## 2 Diffusion approximation and fixation probability

Equation 27 gives the expected change in the frequency of strategy $C$. We turn to calculate the variance of $\Delta p_C$. We have

$$\begin{aligned} \mathrm{Var}(\Delta p_C) &= \mathbb{E}\left(\Delta p_C^2\right) - (\mathbb{E}(\Delta p_C))^2 \\ &= \frac{1}{N^2} \left(\mathbb{P}\left(\Delta p_C = \frac{1}{N}\right) + \mathbb{P}\left(\Delta p_C = -\frac{1}{N}\right)\right)\Big|_{\theta=0} + O(\theta). \\ &= \frac{2}{N^2(\ell-1)} \frac{p_C(1-p_C)[(\ell-1)k - \ell]}{k-1} \end{aligned} \tag{28}$$

We apply Dynkin's Formula to calculate the fixation probability of strategy $C$, the probability that the population reaches the all-$C$ absorbing state starting from the initial condition $p_C(t=0) = p_0$. Specifically, the fixation probability $\rho(p_0)$ satisfies the following differential equation

$$\mathbb{E}(\Delta p_C)|_{p_C=p_0} \frac{d\rho(p_0)}{dp_0} + \frac{\mathrm{Var}(\Delta p_C)|_{p_C=p_0}}{2} \frac{d^2\rho(p_0)}{dp_0^2} = 0. \tag{29}$$

To solve Eq. 29, we introduce the quantity

$$G(p_C) = \exp\left(-\int \frac{2\mathbb{E}(\Delta p_C)}{\mathrm{Var}(\Delta p_C)}\, dp_C\right). \tag{30}$$

We notice that, for any $n_D \geq 1$, the quantity $q_{(n_C,n_D)|C}$ has a factor $1-p_C$. It follows that

$$\frac{q_{(n_C,n_D)|C}}{1-p_C} = \sum_{m=0}^{\ell-2} \gamma_{n_C,n_D}^{(m)} p_C^m, \tag{31}$$

where $\gamma_{n_C,n_D}^{(m)} = \sum_{i=0}^{m} \alpha_{n_C,n_D}^{(i)}$. Then, by Eq. 22, we obtain that the ratio between $\mathbb{E}(\Delta p_C)$ and $\mathrm{Var}(\Delta p_C)$ can also be expressed as a polynomial function of $p_C$,

$$\frac{2\mathbb{E}(\Delta p_C)}{\mathrm{Var}(\Delta p_C)} = \theta \sum_{m=0}^{\ell-1} \overline{\tau}^{(m)} p_C^m, \tag{32}$$

where

$$\begin{aligned}
\mu^{(m)} &:= \sum_{n_C+n_D=\ell-1} \left(\partial_\theta F_C\left(n_C+1, n_D-1; 0\right) \frac{n_D}{n_C+1} - \partial_\theta G_D\left(n_C, n_D; 0\right)\right) \gamma_{n_C,n_D}^{(m)} \quad (0 \leq m \leq \ell-2),\ \mu^{(\ell-1)} := 0, \\
\xi^{(m)} &:= \sum_{n_C+n_D=\ell-1} \begin{pmatrix} \partial_\theta F_C(n_C+1,n_D-1;0)+\partial_\theta F_D(n_C+1,n_D-1;0) \\ -\partial_\theta G_C(n_C,n_D;0)-\partial_\theta G_D(n_C,n_D;0) \end{pmatrix} \gamma_{n_C,n_D}^{(m)} \quad (0 \leq m \leq \ell-2),\ \xi^{(\ell-1)} := 0, \\
v^{(m)} &:= \sum_{n_C+n_D=\ell-1} \left(\partial_\theta F_C\left(n_C, n_D; 0\right) + \partial_\theta F_D\left(n_C, n_D; 0\right)\right) \beta_{n_C,n_D}^{(m)} - \left(\partial_\theta G_C\left(n_C, n_D; 0\right) + \partial_\theta G_D\left(n_C, n_D; 0\right)\right) \alpha_{n_C,n_D}^{(m)}, \\
\overline{\mu}^{(m)} &:= \frac{N(k-1)}{(\ell-1)k-\ell} \mu^{(m)}, \\
\overline{\xi}^{(m)} &:= \frac{N(k-1)}{k(\ell-1)\left[(\ell-1)k-\ell\right]} \xi^{(m)}, \\
\overline{v}^{(m)} &:= \frac{N(k-1)}{k(\ell-1)} v^{(m)}, \\
\overline{\tau}^{(m)} &:= \overline{\mu}^{(m)} - \overline{\xi}^{(m)} - \overline{v}^{(m)}.
\end{aligned} \tag{33}$$

Then, the fixation probability of $C$ is given by

$$\rho(p_0) = \frac{\int_0^{p_0} G(p_C)\, dp_C}{\int_0^1 G(p_C)\, dp_C} = p_0 + \theta\left(p_0 \sum_{m=0}^{\ell-1} \frac{1}{(m+1)(m+2)} \overline{\tau}^{(m)} - \sum_{m=0}^{\ell-1} \frac{1}{(m+1)(m+2)} \overline{\tau}^{(m)} p_0^{m+2}\right) + O\left(\theta^2\right). \tag{34}$$

Define $\rho_C := \rho(1/N)$ as the fixation probability of a single $C$-individual in a population of $N-1$ $D$-individuals. In the large-population limit (i.e. $N \gg 1$), the absolute success of cooperation, $\rho_C > 1/N$, is equivalent to

$$\sum_{m=0}^{\ell-1} \frac{1}{(m+1)(m+2)} \overline{\tau}^{(m)} > 0. \tag{35}$$

We can also calculate the fixation probability of a single $D$-individual, $\rho_D$, by the relation $\rho_D = 1 - \rho\,(1 - 1/N)$, and the relative success of cooperation, $\rho_C > \rho_D$, is equivalent to

$$\sum_{m=0}^{\ell-1} \frac{1}{m+1}\overline{\tau}^{(m)} > 0. \tag{36}$$

## 3 Evolutionary game dynamics

The above analytical framework allows us to analyze a broad range of stochastic processes on hypergraphs. We begin with the evolutionary game dynamics on hypergraphs. In this scenario, every individual either chooses cooperation (C) or defection (D) and interacts with their neighbors via hyperedges. The interaction is described by an $\ell$-player game with the most general payoff matrix as shown in Table 1.

**Table 1:** $\ell$-player game on hyperedges

| $n_C$ | 0 | 1 | ... | $\ell-1$ |
|---|---|---|---|---|
| $C$ | $a_0$ | $a_1$ | ... | $a_{\ell-1}$ |
| $D$ | $b_0$ | $b_1$ | ... | $b_{\ell-1}$ |

Here, $a_{n_C}$ (resp. $b_{n_C}$) represents the payoffs of the cooperative (resp. defective) focal player, and $n_C$ represents the number of cooperative neighbors in the hyperedge. For convenience, we additionally define $a_{-1} = b_{-1} = a_\ell = b_\ell = 0$.

We also begin by considering the update of a defector. The payoff of a cooperative neighbor in the hyperedge of type $(D, n_C, n_D)$ is

$$u_C(n_C, n_D) = a_{n_C-1} + (k-1) \sum_{n_C^{(1)}+n_D^{(1)}=\ell-1} a_{n_C^{(1)}} q_{(n_C^{(1)}, n_D^{(1)})|C}, \tag{37}$$

and the payoff of a defective neighbor in this hyperedge is

$$u_D(n_C, n_D) = b_{n_C} + (k-1) \sum_{n_C^{(1)}+n_D^{(1)}=\ell-1} b_{n_C^{(1)}} q_{(n_C^{(1)}, n_D^{(1)})|D}. \tag{38}$$

The fitness of each cooperative and defective neighbor is $f_C(n_C, n_D) = 1 + \theta u_C(n_C, n_D)$ and $f_D(n_C, n_D) = 1 + \theta u_D(n_C, n_D)$ under weak selection.

Likewise, when updating a cooperator, the payoff of a cooperative neighbor in the hyperedge of type $(C, n_C, n_D)$ is

$$v_C(n_C, n_D) = a_{n_C} + (k-1) \sum_{n_C^{(1)}+n_D^{(1)}=\ell-1} a_{n_C^{(1)}} q_{(n_C^{(1)}, n_D^{(1)})|C}, \tag{39}$$

and the payoff of a defective neighbor in this hyperedge is

$$v_D(n_C, n_D) = b_{n_C+1} + (k-1) \sum_{n_C^{(1)}+n_D^{(1)}=\ell-1} b_{n_C^{(1)}} q_{(n_C^{(1)}, n_D^{(1)})|D}. \tag{40}$$

The fitness of each cooperative and defective neighbor is $g_C(n_C, n_D) = 1 + \theta v_C(n_C, n_D)$ and $g_D(n_C, n_D) = 1 + \theta v_D(n_C, n_D)$ under weak selection. In this case,

$$\begin{aligned} F_C(n_C, n_D; \theta) &= n_C f_C(n_C, n_D),\ F_D(n_C, n_D; \theta) = n_D f_D(n_C, n_D), \\ G_C(n_C, n_D; \theta) &= n_C g_C(n_C, n_D),\ G_D(n_C, n_D; \theta) = n_D g_D(n_C, n_D). \end{aligned} \tag{41}$$

and

$$\begin{aligned} \partial_\theta F_C(n_C, n_D; 0) &= n_C u_C(n_C, n_D),\ \partial_\theta G_C(n_C, n_D; 0) = n_C v_C(n_C, n_D), \\ \partial_\theta F_D(n_C, n_D; 0) &= n_D u_D(n_C, n_D),\ \partial_\theta G_D(n_C, n_D; 0) = n_D v_D(n_C, n_D). \end{aligned} \tag{42}$$

Plugging Eq. 42 into Eq. 18 gives

$$\dot{p}_C = \frac{\theta}{N}\left\{\frac{\begin{pmatrix} p_D \sum_{n_C+n_D=\ell-1} q_{(n_C,n_D)|D} n_C u_C(n_C,n_D) \\ -p_C \sum_{n_C+n_D=\ell-1} q_{(n_C,n_D)|C} n_D v_D(n_C,n_D) \end{pmatrix}}{\ell-1} - \frac{1}{k(\ell-1)^2}\left[\begin{pmatrix} p_D \sum_{n_C+n_D=\ell-1} q_{(n_C,n_D)|D} n_C (n_C u_C(n_C,n_D)+n_D u_D(n_C,n_D)) \\ -p_C \sum_{n_C+n_D=\ell-1} q_{(n_C,n_D)|C} n_D (n_C v_C(n_C,n_D)+n_D v_D(n_C,n_D)) \end{pmatrix}\right.\right.$$
$$\left.\left. + \begin{pmatrix} (k-1) p_D z_D \sum_{n_C+n_D=\ell-1} q_{n_C,n_D|D} (n_C u_C(n_C,n_D)+n_D u_D(n_C,n_D)) \\ -(k-1) p_C z_C \sum_{n_C^{(2)}+n_D^{(2)}=\ell-1} q_{n_C,n_D|C} (n_C v_C(n_C,n_D)+n_D v_D(n_C,n_D)) \end{pmatrix}\right]\right\} + O\left(\theta^2\right). \tag{43}$$

Before plugging Eq. 22 into Eq. 43, we first derive useful identities to simplify Eq. 43. By the first $\ell-1$ equations in Eq. 21, we have,

$$\sum_{m=1}^{\ell-1} (\ell-m)\, x_m = \frac{1}{k-1}, \tag{44}$$

and using this identity, we have

$$\begin{aligned} z_C &= \frac{(1-p_C)\,((\ell-1)\,k-\ell)}{k-1},\ \bar{z}_C := \sum_{n_C+n_D=\ell-1} n_C q_{(n_C,n_D)|C} = \ell-1-z_C, \\ z_D &= \frac{p_C\,((\ell-1)\,k-\ell)}{k-1},\ \bar{z}_D := \sum_{n_C+n_D=\ell-1} n_D q_{(n_C,n_D)|D} = \ell-1-z_D. \end{aligned} \tag{45}$$

We define

$$T_C := \sum_{n_C+n_D=\ell-1} q_{(n_C,n_D)|C} a_{n_C},\ T_D := \sum_{n_C+n_D=\ell-1} q_{(n_C,n_D)|D} b_{n_C}. \tag{46}$$

The payoffs in Eqs. 37-40 are simplified as

$$\begin{aligned} u_C(n_C,n_D) &= a_{n_C-1} + (k-1)T_C,\ v_C(n_C,n_D) = a_{n_C} + (k-1)T_C, \\ u_D(n_C,n_D) &= b_{n_C} + (k-1)T_D,\ v_D(n_C,n_D) = b_{n_C+1} + (k-1)T_D. \end{aligned} \tag{47}$$

And we also have the following identities

$$u_C(n_C+1,n_D-1) = v_C(n_C,n_D),\ u_D(n_C+1,n_D-1) = v_D(n_C,n_D). \tag{48}$$

Similar to Eq. 15, we can prove the following identities:

$$\begin{aligned}
p_D \sum_{n_C+n_D=\ell-1} n_C q_{(n_C,n_D)|D} u_C(n_C,n_D) &= p_C \sum_{n_C+n_D=\ell-1} n_D q_{(n_C,n_D)|C} v_C(n_C,n_D), \\
p_D \sum_{n_C+n_D=\ell-1} n_C q_{(n_C,n_D)|D} u_D(n_C,n_D) &= p_C \sum_{n_C+n_D=\ell-1} n_D q_{(n_C,n_D)|C} v_D(n_C,n_D), \\
p_D \sum_{n_C+n_D=\ell-1} n_C n_C q_{(n_C,n_D)|D} &= p_C \sum_{n_C+n_D=\ell-1} n_C n_D q_{(n_C,n_D)|C} + p_C z_C, \\
p_D \sum_{n_C+n_D=\ell-1} n_C n_D q_{(n_C,n_D)|D} &= p_C \sum_{n_C+n_D=\ell-1} n_D n_D q_{(n_C,n_D)|C} - p_C z_C, \\
p_D \sum_{n_C+n_D=\ell-1} n_C n_C q_{(n_C,n_D)|D} u_C(n_C,n_D) &= p_C \sum_{n_C+n_D=\ell-1} n_C n_D q_{(n_C,n_D)|C} v_C(n_C,n_D) \\
&\quad + p_C \sum_{n_C+n_D=\ell-1} n_D q_{(n_C,n_D)|C} v_C(n_C,n_D), \\
p_D \sum_{n_C+n_D=\ell-1} n_C n_D q_{(n_C,n_D)|D} u_D(n_C,n_D) &= p_C \sum_{n_C+n_D=\ell-1} n_D n_D q_{(n_C,n_D)|C} v_D(n_C,n_D) \\
&\quad - p_C \sum_{n_C+n_D=\ell-1} n_D q_{(n_C,n_D)|C} v_D(n_C,n_D), \\
p_D \sum_{n_C+n_D=\ell-1} n_D q_{(n_C,n_D)|D} u_D(n_C,n_D) &= k(\ell-1)(1-p_C)T_D \\
&\quad - p_C \sum_{n_C+n_D=\ell-1} n_D q_{(n_C,n_D)|C} v_D(n_C,n_D), \\
p_C \sum_{n_C+n_D=\ell-1} n_C q_{(n_C,n_D)|C} v_C(n_C,n_D) &= k(\ell-1)p_C T_C \\
&\quad - p_C \sum_{n_C+n_D=\ell-1} n_D q_{(n_C,n_D)|C} v_C(n_C,n_D).
\end{aligned} \tag{49}$$

Substituting Eq. 49 into Eq. 43 gives

$$\mathbb{E}\left(\Delta p_C\right) = \dot{p}_C = \frac{\theta}{N}\left(\frac{p_C \Omega}{k(\ell-1)} - \frac{k-1}{\ell-1} z_C p_C \left(T_D - T_C\right)\right) + O\left(\theta^2\right), \tag{50}$$

where

$$\begin{aligned}
\Omega &= \sum_{n_C+n_D=\ell-1} n_D \left(v_C^{(n_C,n_D)} - v_D^{(n_C,n_D)}\right) q_{(n_C,n_D)|C} \\
&= \sum_{m=0}^{\ell-1} \left(\sum_{n_C+n_D=\ell-1} n_D \left(a_{n_C} - b_{n_C+1}\right) \alpha_{n_C,n_D}^{(m)}\right) p_C^m + (k-1)\left(T_C - T_D\right) z_C
\end{aligned}. \tag{51}$$

In this case, we have

$$\begin{aligned}
\mu^{(m)} &:= \sum_{n_C+n_D=\ell-1, n_D\geq 1} n_D \left(a_{n_C} - b_{n_C+1}\right) \gamma_{n_C,n_D}^{(m)} \quad (0 \leq m \leq \ell-2), \ \mu^{(\ell-1)} := 0, \\
\nu^{(m)} &:= \sum_{n_C+n_D=\ell-1} b_{n_C} \beta_{n_C,n_D}^{(m)} - a_{n_C} \alpha_{n_C,n_D}^{(m)} \quad (0 \leq m \leq \ell-1), \\
\overline{\mu}^{(m)} &:= \frac{N(k-1)}{k\left[(\ell-1)k - \ell\right]} \mu^{(m)}, \\
\overline{\nu}^{(m)} &:= \frac{N(k+1)(k-1)}{k} \nu^{(m)}, \\
\overline{\tau}^{(m)} &:= \overline{\mu}^{(m)} - \overline{\nu}^{(m)}.
\end{aligned} \tag{52}$$

Plugging Eq. 52 into Eq. 34 gives the fixation probability of cooperation, and the condition for $\rho_C > 1/N$ (Eq. 35) becomes

$$\sum_{m=0}^{\ell-1} \frac{1}{(m+1)(m+2)} \left(\overline{\mu}^{(m)} - \overline{\nu}^{(m)}\right) > 0. \tag{53}$$

### 3.1 Well-mixed populations

In a well-mixed population, the average degree $k$ equals $\binom{N-1}{\ell-1}$, which is much greater than 1 when $N \gg 1$. As a result, we only need to consider the terms of the highest order in $k$. We notice that the coefficients of $\overline{v}^{(m)}$ and $\overline{u}^{(m)}$ in Eq. 52 are $O(k)$ and $O(1/k)$. Then $\overline{u}^{(m)}$ in Eq. 53 is ignored, and Eq. 53 is equivalent to

$$\sum_{m=0}^{\ell-1} \frac{1}{(m+1)(m+2)} v^{(m)} < 0. \tag{54}$$

We then move to calculate the coefficient of $\alpha^{(m)}_{n_C,n_D}$ and $\beta^{(m)}_{n_C,n_D}$, keeping only the highest-order terms in $k$. By Eqs. 24-26, the coefficient of $\alpha^{(m)}_{n_C,n_D}$ is proportional to

$$\varphi_{n_C,n_D} = (-1)^{n_D} \binom{\ell-1}{n_D} \sum_{i=0}^{n_D} (-1)^{n_D-i} \binom{n_D}{i} \left( \frac{1}{i+n_C+1} - \frac{1}{i+n_C+2} \right). \tag{55}$$

Since

$$\begin{aligned} \sum_{i=0}^{\ell} \frac{1}{i+m} \binom{\ell}{i} (-1)^{l-i} &= \sum_{i=0}^{\ell} \left( \int_0^1 x^{i+m-1}\, dx \right) \binom{\ell}{i} (-1)^{\ell-i} = \int_0^1 x^{m-1} \left( \sum_{i=0}^{\ell} \binom{\ell}{i} (-1)^{\ell-i} x^i \right) dx \\ &= (-1)^{\ell} \int_0^1 x^{m-1} (1-x)^{\ell}\, dx = (-1)^{\ell} \frac{(m-1)!\ell!}{(m+\ell)!}, \end{aligned} \tag{56}$$

we have

$$\varphi_{n_C,n_D} = \binom{\ell-1}{n_D} \left( \frac{n_C! n_D!}{\ell!} - \frac{(n_C+1)! n_D!}{(\ell+1)!} \right) = \frac{n_D+1}{\ell(\ell+1)}. \tag{57}$$

Similarly, we can prove that the coefficient of $\beta^{(m)}_{n_C,n_D}$ is also proportional to $(n_D+1)/(\ell(\ell+1))$. Then, $\rho_C > 1/N$ holds when

$$\sum_{n_C=0}^{\ell-1} (\ell-n_C) a_{n_C} > \sum_{n_C=0}^{\ell-1} (\ell-n_C) b_{n_C}, \tag{58}$$

which recovers the results in Ref. (2)

### 3.2 $k$-regular graphs

In this section, we provide explicit results for $\ell = 2$ and $\ell = 3$ on hypergraphs of average degree $k$. For $\ell = 2$, we have

$$\begin{aligned} &\mu^{(0)} = (a_0 - b_1)\, \alpha^{(0)}_{0,1},\ \mu^{(1)} = 0,\ v^{(m)} = b_0 \beta^{(m)}_{0,1} + b_1 \beta^{(m)}_{1,0} - a_0 \alpha^{(m)}_{0,1} - a_1 \alpha^{(m)}_{1,0} \quad (m = 0, 1), \\ &\overline{\mu}^{(m)} = \frac{N(k-1)}{k(k-2)} \mu^{(m)},\ \overline{v}^{(m)} = \frac{N(k+1)(k-1)}{k} v^{(m)}, \\ &\alpha^{(0)}_{1,0} = \frac{1}{k-1},\ \alpha^{(1)}_{1,0} = \frac{k-2}{k-1},\ \alpha^{(0)}_{0,1} = \frac{k-2}{k-1},\ \alpha^{(1)}_{0,1} = -\frac{k-2}{k-1}, \\ &\beta^{(0)}_{1,0} = 0,\ \beta^{(1)}_{1,0} = \frac{k-2}{k-1},\ \beta^{(0)}_{0,1} = 1,\ \beta^{(1)}_{0,1} = -\frac{k-2}{k-1}. \end{aligned} \tag{59}$$

Plugging Eq. 59 into Eq. 53, we have

$$(2k^2 - 2k - 1)a_0 + (k^2 + 2k + 1)a_1 > (2k^2 + k - 1)b_0 + (k^2 - k + 1)b_1, \tag{60}$$

which recovers the result in pairwise regular graphs, derived in Ref. (1).

For $\ell = 3$, we have

$$
\begin{aligned}
&\mu^{(0)} = (a_1 - b_2)\,\gamma_{1,1}^{(0)} + 2\,(a_0 - b_1)\,\gamma_{0,2}^{(0)},\ \mu^{(1)} = (a_1 - b_2)\,\gamma_{1,1}^{(1)} + 2\,(a_0 - b_1)\,\gamma_{0,2}^{(1)},\ \mu^{(2)} = 0,\\
&\gamma_{1,1}^{(0)} = \alpha_{1,1}^{(0)},\ \gamma_{0,2}^{(0)} = \alpha_{0,2}^{(0)},\ \gamma_{1,1}^{(1)} = \alpha_{1,1}^{(1)} + \alpha_{1,1}^{(0)},\ \gamma_{0,2}^{(1)} = \alpha_{0,2}^{(1)} + \alpha_{0,2}^{(0)},\\
&v^{(m)} = b_0\beta_{0,2}^{(m)} + b_1\beta_{1,1}^{(m)} + b_2\beta_{2,0}^{(m)} - a_0\alpha_{0,2}^{(m)} - a_1\alpha_{1,1}^{(m)} - a_2\alpha_{2,0}^{(m)} \quad (m = 0,1,2),\\
&\overline{\mu}^{(m)} = \frac{N(k-1)}{k(2k-3)}\mu^{(m)},\ \overline{v}^{(m)} = \frac{N(k+1)(k-1)}{k}v^{(m)},
\end{aligned}
\tag{61}
$$

and

$$
\begin{aligned}
&\alpha_{2,0}^{(0)} = \frac{1}{(2k-1)(k-1)},\ \alpha_{2,0}^{(1)} = \frac{6k-9}{2(2k-1)(k-1)},\ \alpha_{2,0}^{(2)} = \frac{(2k-3)^2}{2(2k-1)(k-1)},\\
&\alpha_{1,1}^{(0)} = \frac{2k-3}{(2k-1)(k-1)},\ \alpha_{1,1}^{(1)} = \frac{(2k-3)(2k-4)}{(2k-1)(k-1)},\ \alpha_{1,1}^{(2)} = -\frac{(2k-3)^2}{(2k-1)(k-1)},\\
&\alpha_{0,2}^{(0)} = \frac{(k-1)(2k-3)}{(2k-1)(k-1)},\ \alpha_{0,2}^{(1)} = -\frac{(2k-3)(4k-5)}{2(2k-1)(k-1)},\ \alpha_{0,2}^{(2)} = \frac{(2k-3)^2}{2(2k-1)(k-1)},\\
&\beta_{2,0}^{(0)} = 0,\ \beta_{2,0}^{(1)} = \frac{2k-3}{2(2k-1)(k-1)},\ \beta_{2,0}^{(2)} = \frac{(2k-3)^2}{2(2k-1)(k-1)},\\
&\beta_{1,1}^{(0)} = 0,\ \beta_{1,1}^{(1)} = \frac{(2k-2)(2k-3)}{(2k-1)(k-1)},\ \beta_{1,1}^{(2)} = -\frac{(2k-3)^2}{(2k-1)(k-1)},\\
&\beta_{0,2}^{(0)} = 1,\ \beta_{0,2}^{(1)} = -\frac{(2k-3)(4k-3)}{2(2k-1)(k-1)},\ \beta_{0,2}^{(2)} = \frac{(2k-3)^2}{2(2k-1)(k-1)}.
\end{aligned}
\tag{62}
$$

In this case, $\rho_C > 1/N$ is equivalent to

$$
\begin{aligned}
&(12k^3 - 16k^2 + 3k + 3)a_0 + (8k^3 - 6k - 6)a_1 + (4k^3 + 4k^2 + 3k + 3)a_2\\
&> (12k^3 - 9k + 3)b_0 + (8k^3 - 8k^2 + 6k - 6)b_1 + (4k^3 - 4k^2 + 3k + 3)b_2
\end{aligned}.
\tag{63}
$$

We then consider two specific types of multiplayer social dilemma: the donation game and the nonlinear public goods game.

**Donation game (DG).** The two-player donation game has the following payoff matrix:

$$
\begin{pmatrix} a_0 & a_1 \\ b_0 & b_1 \end{pmatrix} = \begin{pmatrix} -c & -c+b \\ 0 & b \end{pmatrix}.
\tag{64}
$$

Plugging these elements into Eq. 60 recovers the well-known $b/c > k$ condition for cooperation to be favored.

To investigate the structure of the multi-player donation game, which leads to the $b/c > k$ condition, we begin by analyzing the case of $\ell = 3$ with Eq. 63. We notice that when the three-player game has the following payoff matrix:

$$
\begin{pmatrix} a_0 & a_1 & a_2 \\ b_0 & b_1 & b_2 \end{pmatrix} = \begin{pmatrix} -c & -c+b & -c+2b \\ 0 & b & 2b \end{pmatrix},
\tag{65}
$$

Eq. 63 is equivalent to $b/c > k$.

Based on this observation, we turn to prove that, for any $\ell$, the $b/c > k$ condition holds when

$$
\begin{pmatrix} a_0 & a_1 & \dots & a_{\ell-1} \\ b_0 & b_1 & \dots & b_{\ell-1} \end{pmatrix} = \begin{pmatrix} -c & -c+b & \dots & -c+(\ell-1)b \\ 0 & b & \dots & (\ell-1)b \end{pmatrix}.
\tag{66}
$$

In this case, instead of calculating every $\alpha_{n_C,n_D}^{(m)}$ and $\beta_{n_C,n_D}^{(m)}$, we directly calculate the quantities $\Omega$, $T_C$, and $T_D$. We

have

$$\begin{aligned}
T_C &= -c + b\left(\ell - 1 - \frac{(1-p_C)\left((\ell-1)k - \ell\right)}{k-1}\right), \\
T_D &= b\frac{p_C\left((\ell-1)k-\ell\right)}{k-1}, \\
\Omega &= \frac{(1-p_C)\left((\ell-1)k-\ell\right)}{k-1}\left((k-1)\left(T_C - T_D\right) - (b+c)\right).
\end{aligned} \tag{67}$$

Plugging these quantities into Eq. 32 gives

$$\frac{2\mathbb{E}(\Delta p_C)}{\mathrm{Var}(\Delta p_C)} = \theta N\left(b - kc\right), \tag{68}$$

which means

$$\overline{\tau}^{(0)} = N\left(b-kc\right),\ \overline{\tau}^{(m)} = 0 \quad \left(1 \le m \le \ell - 1\right). \tag{69}$$

Then, we obtain the $b/c > k$ condition by Eq. 53.

**Public goods game (PGG).** For the two-player public goods game

$$\begin{pmatrix} a_0 & a_1 \\ b_0 & b_1 \end{pmatrix} = \begin{pmatrix} -c + \dfrac{b}{2} & -c + \dfrac{(\delta_2+1)b}{2} \\ 0 & \dfrac{b}{2} \end{pmatrix}, \tag{70}$$

and the three-player public goods game

$$\begin{pmatrix} a_0 & a_1 & a_2 \\ b_0 & b_1 & b_2 \end{pmatrix} = \begin{pmatrix} -c + \dfrac{b}{3} & -c + \dfrac{(\delta_3+1)b}{3} & -c + \dfrac{(\delta_3^2+\delta_3+1)b}{3} \\ 0 & \dfrac{b}{3} & \dfrac{(\delta_3+1)b}{3} \end{pmatrix}, \tag{71}$$

we have

$$\rho_C(1/N) > \frac{1}{N} \Leftrightarrow \frac{b}{c} > \left(\frac{b}{c}\right)^* = \begin{cases} \dfrac{6k^2}{(k+1)(k(\delta_2+2)+\delta_2-1)} & \ell = 2, \\ \dfrac{36k^2(2k-1)}{(k+1)\left(\delta_3^2\left(4k^2+3\right)+\delta_3\left(8k^2-6\right)+3(2k-1)^2\right)} & \ell = 3. \end{cases} \tag{72}$$

## 4 Neutral complex contagion

The complexity of higher-order interactions in evolutionary dynamics arises from game interactions. In this section, we move to analyze neutral complex contagion on hypergraphs, where this complexity is at work during the strategy updating process. In this case, strategies $C$ and $D$ can be interpreted as two competing opinions. The term "neutral" here indicates that neither opinion has an evolutionary advantage over the other. More specifically, we have

$$\begin{aligned}
F_C(n_C, n_D; \theta) &= G_C(n_C, n_D; \theta) = n_C^{1+\lambda\theta}, \\
F_D(n_C, n_D; \theta) &= G_D(n_C, n_D; \theta) = n_D^{1+\lambda\theta},
\end{aligned} \tag{73}$$

where $\lambda$ represents the complexity parameter in contagion dynamics. When $\lambda = 0$, opinions spread via simple contagion. Otherwise, opinions spread via complex contagion. We have

$$\begin{aligned}
\partial_\theta F_C(n_C, n_D; 0) &= \partial_\theta G_C(n_C, n_D; 0) = \lambda n_C \ln n_C, \\
\partial_\theta F_D(n_C, n_D; 0) &= \partial_\theta G_D(n_C, n_D; 0) = \lambda n_D \ln n_D.
\end{aligned} \tag{74}$$

Similar to Eq. 49, in this case, we can prove the following identities

$$\begin{aligned} p_D \sum_{n_C+n_D=\ell-1} q_{(n_C,n_D)|D} n_C \ln n_C &= p_C \sum_{n_C+n_D=\ell-1} q_{(n_C,n_D)|C} n_D \ln(n_C+1), \\ p_D \sum_{n_C+n_D=\ell-1} q_{(n_C,n_D)|D} n_C n_C \ln n_C &= p_C \sum_{n_C+n_D=\ell-1} q_{(n_C,n_D)|C} n_D n_C \ln(n_C+1) \\ &\quad + p_C \sum_{n_C+n_D=\ell-1} q_{(n_C,n_D)|C} n_D \ln(n_C+1), \\ p_D \sum_{n_C+n_D=\ell-1} q_{(n_C,n_D)|D} n_C n_D \ln n_D &= p_C \sum_{n_C+n_D=\ell-1, n_D\geq 2} q_{(n_C,n_D)|C} n_D n_D \ln(n_D-1) \\ &\quad - p_C \sum_{n_C+n_D=\ell-1, n_D\geq 2} q_{(n_C,n_D)|C} n_D \ln(n_D-1). \end{aligned} \tag{75}$$

As a result, the time derivative of $p_C$ becomes

$$\begin{aligned} \dot{p}_C = \mathbb{E}(\Delta p_C) = \frac{\lambda\theta}{N} \Bigg\{ & \frac{p_C \sum_{n_C+n_D=\ell-1} q_{(n_C,n_D)|C} n_D \ln \frac{n_C+1}{n_D}}{\ell-1} - \frac{1}{k(\ell-1)^2} \Bigg[ \begin{pmatrix} p_C \sum_{n_C+n_D=\ell-1} q_{(n_C,n_D)|C} n_C n_D \ln\left(\frac{n_C+1}{n_C}\right) \\ + p_C \sum_{n_C+n_D=\ell-1} q_{(n_C,n_D)|C} n_D \ln(n_C+1) \end{pmatrix} \\ & + \begin{pmatrix} p_C \sum_{n_C+n_D=\ell-1, n_D\geq 2} q_{(n_C,n_D)|C} n_D n_D \ln(n_D-1) \\ - p_C \sum_{n_C+n_D=\ell-1, n_D\geq 2} q_{(n_C,n_D)|C} n_D \ln(n_D-1) \\ - p_C \sum_{n_C+n_D=\ell-1} q_{(n_C,n_D)|C} n_D n_D \ln n_D \end{pmatrix} + \begin{pmatrix} (k-1) p_D z_D \sum_{n_C+n_D=\ell-1} q_{(n_C,n_D)|D} (n_C \ln n_C + n_D \ln n_D) \\ -(k-1) p_C z_C \sum_{n_C+n_D=\ell-1} q_{(n_C,n_D)|C} (n_C \ln n_C + n_D \ln n_D) \end{pmatrix} \Bigg] \Bigg\} + O\left(\theta^2\right). \end{aligned} \tag{76}$$

The variance of $\Delta p_C$ remains identical to Eq. 28.

We define

$$\begin{aligned} \mu^{(m)} &:= \sum_{n_C+n_D=\ell-1} n_D \ln \frac{n_C+1}{n_D} \gamma^{(m)}_{n_C,n_D} \quad (0 \leq m \leq \ell-2),\ \mu^{(\ell-1)} := 0, \\ \xi_1^{(m)} &:= \sum_{n_C+n_D=\ell-1} n_C n_D \ln \frac{n_C+1}{n_C} \gamma^{(m)}_{n_C,n_D} \quad (0 \leq m \leq \ell-2),\ \xi_1^{(\ell-1)} := 0, \\ \xi_2^{(m)} &:= \sum_{n_C+n_D=\ell-1} n_D \ln(n_C+1)\, \gamma^{(m)}_{n_C,n_D} \quad (0 \leq m \leq \ell-2),\ \xi_2^{(\ell-1)} := 0, \\ \xi_3^{(m)} &:= \sum_{n_C+n_D=\ell-1} n_D n_D \ln n_D \gamma^{(m)}_{n_C,n_D} \quad (0 \leq m \leq \ell-2),\ \xi_3^{(\ell-1)} := 0, \\ \xi_4^{(m)} &:= \sum_{n_C+n_D=\ell-1, n_D\geq 2} n_D n_D \ln(n_D-1)\, \gamma^{(m)}_{n_C,n_D} \quad (0 \leq m \leq \ell-2),\ \xi_4^{(\ell-1)} := 0, \\ \xi_5^{(m)} &:= \sum_{n_C+n_D=\ell-1, n_D\geq 2} n_D \ln(n_D-1)\, \gamma^{(m)}_{n_C,n_D} \quad (0 \leq m \leq \ell-2),\ \xi_5^{(\ell-1)} := 0, \\ v^{(m)} &:= \sum_{n_C+n_D=\ell-1} (n_C \ln n_C + n_D \ln n_D) \left(\beta^{(m)}_{n_C,n_D} - \alpha^{(m)}_{n_C,n_D}\right), \\ \overline{\mu}^{(m)} &:= \frac{\lambda N(k-1)}{(\ell-1)k-\ell} \mu^{(m)}, \\ \overline{\xi}_i^{(m)} &:= \frac{\lambda N(k-1)}{k(\ell-1)\,[(\ell-1)k-\ell]} \xi_i^{(m)} \quad (i = 1,2,3,4,5), \\ \overline{v}^{(m)} &:= \frac{\lambda N(k-1)}{k(\ell-1)} v^{(m)}, \\ \overline{\tau}^{(m)} &:= \overline{\mu}^{(m)} - \overline{\xi}_1^{(m)} - \overline{\xi}_2^{(m)} + \overline{\xi}_3^{(m)} - \overline{\xi}_4^{(m)} + \overline{\xi}_5^{(m)} - \overline{v}^{(m)}. \end{aligned} \tag{77}$$

In addition, to ensure these definitions are all well-defined, when $n_C = 0$, we set $n_C \ln n_C = \lim_{n_C\to 0} n_C \ln n_C = 0$ and $n_C \ln(1/n_C) = \lim_{n_C\to 0} n_C \ln(1/n_C) = 0$. Plugging Eq. 77 into Eq. 34 gives the fixation probability in the regime of neutral complex contagion.

For pairwise graphs (i.e. $\ell = 2$), there is no complexity in contagion dynamics, so the evolutionary process behaves as a martingale, and the fixation probability of opinion $C$ is equal to its initial frequency $p_0$. However,

when the contagion occurs on hypergraphs, strategy $C$ spreads via complex contagion. As an illustrative example, we calculate the fixation probability of $C$ when $\ell = 3$. We have

$$\begin{aligned}
&\mu^{(0)} = \ln 2 \cdot \gamma_{1,1}^{(0)} - 2\ln 2 \cdot \gamma_{0,2}^{(0)},\ \mu^{(1)} = \ln 2 \cdot \gamma_{1,1}^{(1)} - 2\ln 2 \cdot \gamma_{0,2}^{(1)},\ \mu^{(2)} = 0,\\
&\xi_1^{(0)} = \xi_2^{(0)} = \ln 2 \cdot \gamma_{1,1}^{(0)},\ \xi_1^{(1)} = \xi_2^{(1)} = \ln 2 \cdot \gamma_{1,1}^{(1)}, \xi_1^{(2)} = \xi_2^{(2)} = 0,\\
&\xi_3^{(0)} = 4\ln 2 \cdot \gamma_{0,2}^{(0)},\ \xi_3^{(1)} = 4\ln 2 \gamma_{0,2}^{(1)},\ \xi_3^{(2)} = 0,\\
&\gamma_{1,1}^{(0)} = \alpha_{1,1}^{(0)},\ \gamma_{0,2}^{(0)} = \alpha_{0,2}^{(0)},\ \gamma_{1,1}^{(1)} = \alpha_{1,1}^{(0)} + \alpha_{1,1}^{(1)},\ \gamma_{0,2}^{(1)} = \alpha_{0,2}^{(0)} + \alpha_{0,2}^{(1)},\\
&\varphi_1^{(m)} = \varphi_2^{(m)} = 0 \quad (m = 0, 1, 2)\,,\\
&v^{(m)} = 2\ln 2 \left(\beta_{0,2}^{(m)} + \beta_{2,0}^{(m)} - \alpha_{0,2}^{(m)} - \alpha_{2,0}^{(m)}\right) \quad (m = 0, 1, 2)\,,\\
&\overline{\mu}^{(m)} = \frac{\lambda N(k-1)}{2k-3}\mu^{(m)},\ \overline{\xi}_i^{(m)} = \frac{\lambda N(k-1)}{2k\,(2k-3)}\xi_i^{(m)},\ \overline{v}^{(m)} = \frac{\lambda N(k-1)}{2k}v^{(m)}.
\end{aligned} \tag{78}$$

In this case,

$$\rho(p_0) = p_0 + \theta N \lambda \ln 2 \cdot \frac{2k-3}{6(2k-1)} p_0(1-p_0)(2p_0-1) + O\left(\theta^2\right). \tag{79}$$

and

$$\rho_C = \frac{1}{N} + \theta \lambda \ln 2 \cdot \frac{2k-3}{2k-1}\left(-\frac{1}{6} + \frac{1}{2N} - \frac{1}{3N^2}\right) + O\left(\theta^2\right). \tag{80}$$

In the large population limit $N \to \infty$,

$$\rho_C = \frac{1}{N} - \frac{\theta \lambda \ln 2 \cdot (2k-3)}{6(2k-1)} + O\left(\theta^2\right). \tag{81}$$

Since the contagion dynamics are symmetric with respect to both strategies, the fixation probability of $D$, $\rho_D$, is identical to $\rho_C$. In well-mixed populations, the first-order term of $\rho_C$ approaches $-\lambda \ln 2/6$.

## 5 Complex contagion with selection

In evolutionary game dynamics, higher-order interactions govern an individual's fitness and strategic updates, but the strategy updating process remains pairwise-based: each individual copies the strategy of another individual. Conversely, in a neutral complex contagion, all individuals have the same fitness, but the strategy updating process is governed by higher-order interactions: an individual inherits a strategy from its interaction neighborhood rather than from another individual. Therefore, a straightforward extension is to simultaneously consider higher-order interactions in both the assignment of fitness and in the process of strategic updating.

We analyze the following combination:

$$\begin{aligned}
F_C(n_C, n_D; \theta) = n_C^{1+\lambda\theta} f_C(n_C, n_D),\ G_C(n_C, n_D; \theta) = n_C^{1+\lambda\theta} g_C(n_C, n_D),\\
F_D(n_C, n_D; \theta) = n_D^{1+\lambda\theta} f_D(n_C, n_D), G_D(n_C, n_D; \theta) = n_D^{1+\lambda\theta} g_D(n_C, n_D).
\end{aligned} \tag{82}$$

The derivatives with respect to $\theta$ at $\theta = 0$ are

$$\begin{aligned}
\partial_\theta F_C(n_C, n_D; 0) = n_C u_C(n_C, n_D) + \lambda n_C \ln n_C,\ \partial_\theta G_C(n_C, n_D; 0) = n_C v_C(n_C, n_D) + \lambda n_C \ln n_C,\\
\partial_\theta F_D(n_C, n_D; 0) = n_D u_D(n_C, n_D) + \lambda n_D \ln n_D,\ \partial_\theta G_D(n_C, n_D; 0) = n_D v_D(n_C, n_D) + \lambda n_D \ln n_D,
\end{aligned} \tag{83}$$

which are the sum of the derivatives in the cases of evolutionary game dynamics (Eq. 42) and complex contagion (Eq. 74). Then the expectation $\mathbb{E}(\Delta p_C)$ in this case is also the sum of Eq. 50 and Eq. 76.

We are interested in how complex contagion affects the evolution of cooperation on hypergraphs. Specifically, we analyze the effect of the complexity parameter $\lambda$ on the critical benefit-to-cost ratio $(b/c)^*$. We first derive the

fixation probability $\rho_C$ under a general three-player game. For a large population size $N$, we have

$$\rho_C = \frac{1}{N} + \frac{\theta}{24k(2k-1)} \begin{pmatrix} a_0\left(12k^3-16k^2+3k+3\right)+a_1\left(8k^3-6k-6\right)+a_2\left(4k^3+4k^2+3k+3\right) \\ -b_0\left(12k^3-9k+3\right)-b_1\left(8k^3-8k^2+6k-6\right)-b_2\left(4k^3-4k^2+3k+3\right) \end{pmatrix} - 4\lambda k \cdot \ln 2\,(2k-3)\Bigg) + O\left(\theta^2\right). \tag{84}$$

For the three-player donation game (Eq. 65), the fixation probability is

$$\rho_C = \frac{1}{N} + \frac{\theta}{24(2k-1)} \left(12(2k-1)(b-ck) - 4\lambda \cdot \ln 2(2k-3)\right) + O\left(\theta^2\right). \tag{85}$$

For the three-player public goods game (Eq. 71), the fixation probability is

$$\rho_C = \frac{1}{N} + \frac{\theta}{72k(2k-1)} \begin{pmatrix} (k+1)\left(\delta_3^2\left(4k^2+3\right)+\delta_3\left(8k^2-6\right)+3(2k-1)^2\right) \\ -3k(12k(2k-1)c+4\lambda\cdot\ln 2(2k-3)) \end{pmatrix} + O\left(\theta^2\right). \tag{86}$$

When $c = 1$, the critical benefit-to-cost ratio in these two games becomes

$$\left(\frac{b}{c}\right)^* = \begin{cases} k + \dfrac{\lambda \ln 2 \cdot (2k-3)}{3(2k-1)} & \text{Donation game,} \\ \dfrac{36k^2(2k-1) + 12\lambda k \cdot \ln 2(2k-3)}{(k+1)\left(\delta_3^2\left(4k^2+3\right) + \delta_3\left(8k^2-6\right) + 3(2k-1)^2\right)} & \text{Public goods game.} \end{cases} \tag{87}$$

When the evolutionary game dynamics are driven by the average payoffs (i.e., $u$ and $v$ are replaced by $u/k$ and $v/k$), the critical ratios become

$$\left(\frac{b}{c}\right)^* = \begin{cases} k + \dfrac{\lambda k \ln 2 \cdot (2k-3)}{3(2k-1)} & \text{Donation game,} \\ \dfrac{36k^2(2k-1) + 12\lambda k^2 \cdot \ln 2(2k-3)}{(k+1)\left(\delta_3^2\left(4k^2+3\right) + \delta_3\left(8k^2-6\right) + 3(2k-1)^2\right)} & \text{Public goods game.} \end{cases} \tag{88}$$

## 6 Generation of random graphs

In this work, beyond the vertex-transitive graph considered in the main text, we also investigate the performance of diffusion approximation on third-order random graphs, including both regular graphs and heterogeneous graphs generated by the preferential attachment rule, which was introduced in the Barabási-Albert model (3).

A third-order random regular graph is generated by the configuration model outlined in Algorithm 1. The basic idea is to assign each node $k$ stubs and then randomly shuffle and group these stubs into triples, each forming a hyperedge. The same idea can be extended to generate $\ell$th-order random regular graphs.

**Algorithm 1** Configuration model for third-order random regular graphs

**Input:** Number of nodes $N$, average degree $k$, maximum retries $T$
**Output:** Set of hyperedges
Check if $3 \mid Nk$. If not, return "No valid hypergraph"
Set $M \leftarrow Nk/3$: the total number of hyperedges
**for** $t = 1$ to $T$ **do**
    Randomly permute stubs
    Partition stubs into $M$ groups of size 3
    Let each group be a candidate hyperedge
    **if** every hyperedge has 3 distinct vertices **then**
        **if** no duplicate hyperedges **then**
            Count vertex occurrences across hyperedges
            **if** each vertex appears exactly $k$ times and the graph is connected **then return** Set of hyperedges
**return** "Construction failed, try again"

For heterogeneous hypergraphs, the construction begins with a complete graph of size $m_0$, that is, the initial hyperedges consist of all possible 3-node combinations. Then, at each time step, add a node and create $m$ new hyperedges that connect the new node to two existing nodes chosen preferentially by hyperdegree. The construction process stops after $T$ time steps. Algorithm 2 outlines the construction procedure, which can also be extended to hypergraphs of arbitrary order $\ell$.

**Algorithm 2** Preferential attachment hypergraph model ($\ell = 3$)

**Input:** Number of time steps $T$, initial graph size $m_0$, linking number $m$
**Output:** Set of hyperedges

```
Initialize hypergraph with m_0 nodes
Form all possible 3-node hyperedges among the m_0 nodes
Set the hyperdegree of each seed node equal to binom(m_0 - 1, 2)
for t = 1 to T do
    Add a new node v with hyperdegree 0
    for j = 1 to m do
        Select an existing node u with probability proportional to its hyperdegree
        Select another distinct node w with probability proportional to its hyperdegree
        Form a new hyperedge (v, u, w)
        Increase hyperdegree of v, u, w by 1
return Set of hyperedges
```

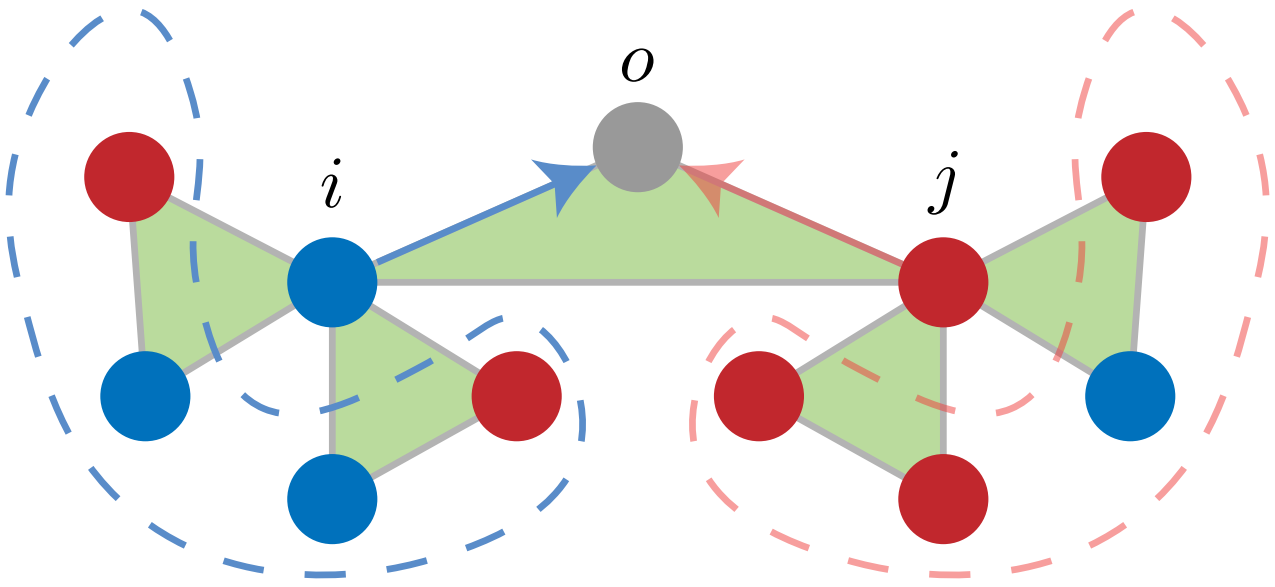


$$N_{C|C} = \sum_{n_C+n_D=\ell-1} n_C q_{(n_C,n_D)|C} \qquad u_i = (k-1)N_{C|C}b - ck$$

$$N_{C|D} = \sum_{n_C+n_D=\ell-1} n_C q_{(n_C,n_D)|D} \qquad u_j = (k-1)N_{C|D}b$$

$$(k-1)\left(N_{C|C} - N_{C|D}\right) = 1$$

$$u_i > u_j \Leftrightarrow \frac{b}{c} > k$$

**Supplementary Figure 1: Intuition for the multiplayer donation game on hypergraphs.** As an illustrative example, we consider the intuition behind the $b/c > k$ rule on hypergraphs of order $\ell = 3$. Suppose an individual $o$ is selected to be replaced, and a cooperator $i$ and a defector $j$ compete for this site. Let $N_{C|C}$ and $N_{C|D}$ denote the average number of cooperative neighbors of a focal cooperation and a focal defector per interaction, respectively. Individual $i$ pays cost $ck$ and receives benefit $(k-1)N_{C|C}b$ from the remaining $(k-1)$ higher-order interactions, while individual $j$ pays no cost but still receives benefit $(k-1)N_{C|D}b$ from cooperative neighbors. Selection favors cooperation when the payoff of $i$, $u_i$, is larger than that of $j$, $u_j$. Since a cooperator has one more cooperative neighbor than a defector on average (Eq. 44), this condition becomes the $b/c > k$ rule.

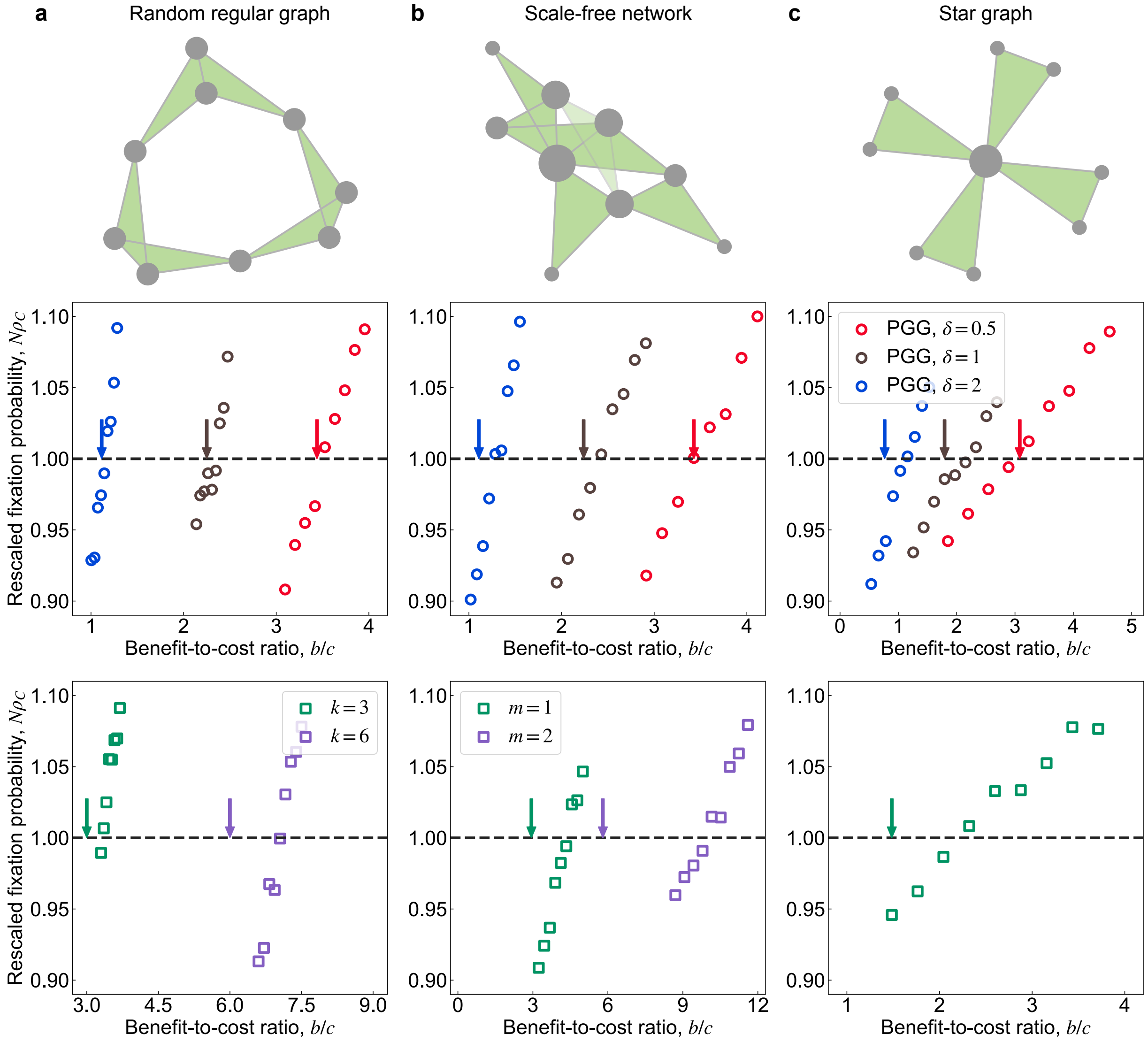


**Supplementary Figure 2: Evolutionary game dynamics on random hypergraphs.** We consider three types of random hypergraphs: random regular hypergraphs with average degree $k$ (**a**) generated by the configuration model (see Algorithm 1); scale-free hypergraphs with linking number $m$ (**b**) generated by the preferential attachment model (see Algorithm 2); higher-order star graphs (**c**) (4) . We calculate the fixation probability of cooperation under the public goods game (the middle panels) and the donation game (the bottom panels), by averaging over $2 \times 10^6$ independent Monte Carlo simulations. Arrows show the theoretical predictions. Although structural heterogeneity causes discrepancies between theory and simulations, the critical benefit-to-cost ratio derived by the $\ell$-hyperedge approximation still provides a necessary condition for the evolution of cooperation. Parameters: $c = 1$, $\theta = 0.01$, $k = 3$ and $m = 1$ in the public goods game.

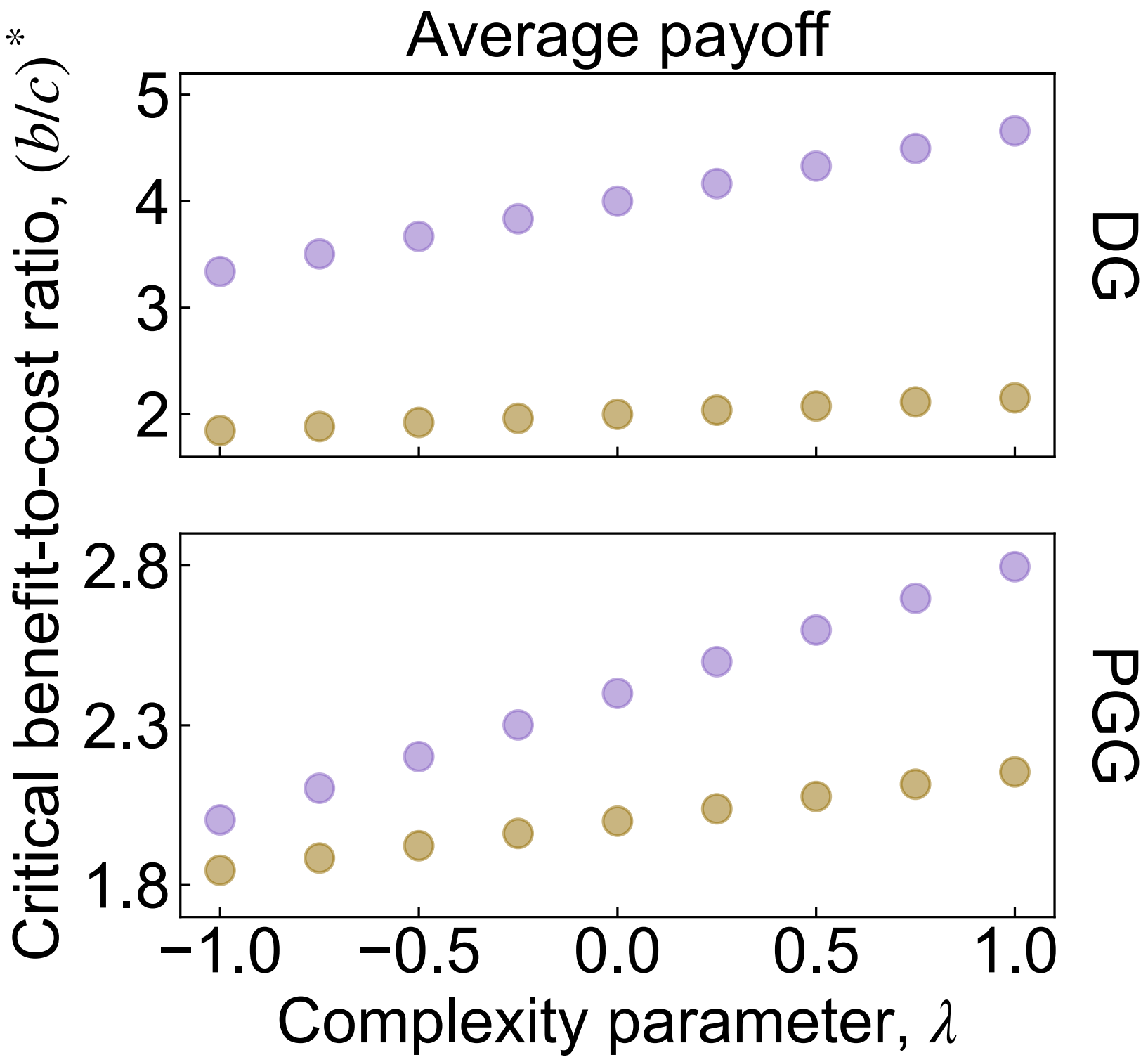


**Supplementary Figure 3: Critical benefit-to-cost ratio with complex contagion under the average-payoff setting.** The effects of the game and complex contagion on the fixation probability $\rho_C$ have an identical order in $k$, then the variants in the critical benefit-to-cost ratio $(b/c)^*$ are more significant as $\lambda$ changes, compared to that under the accumulated-payoff setting.

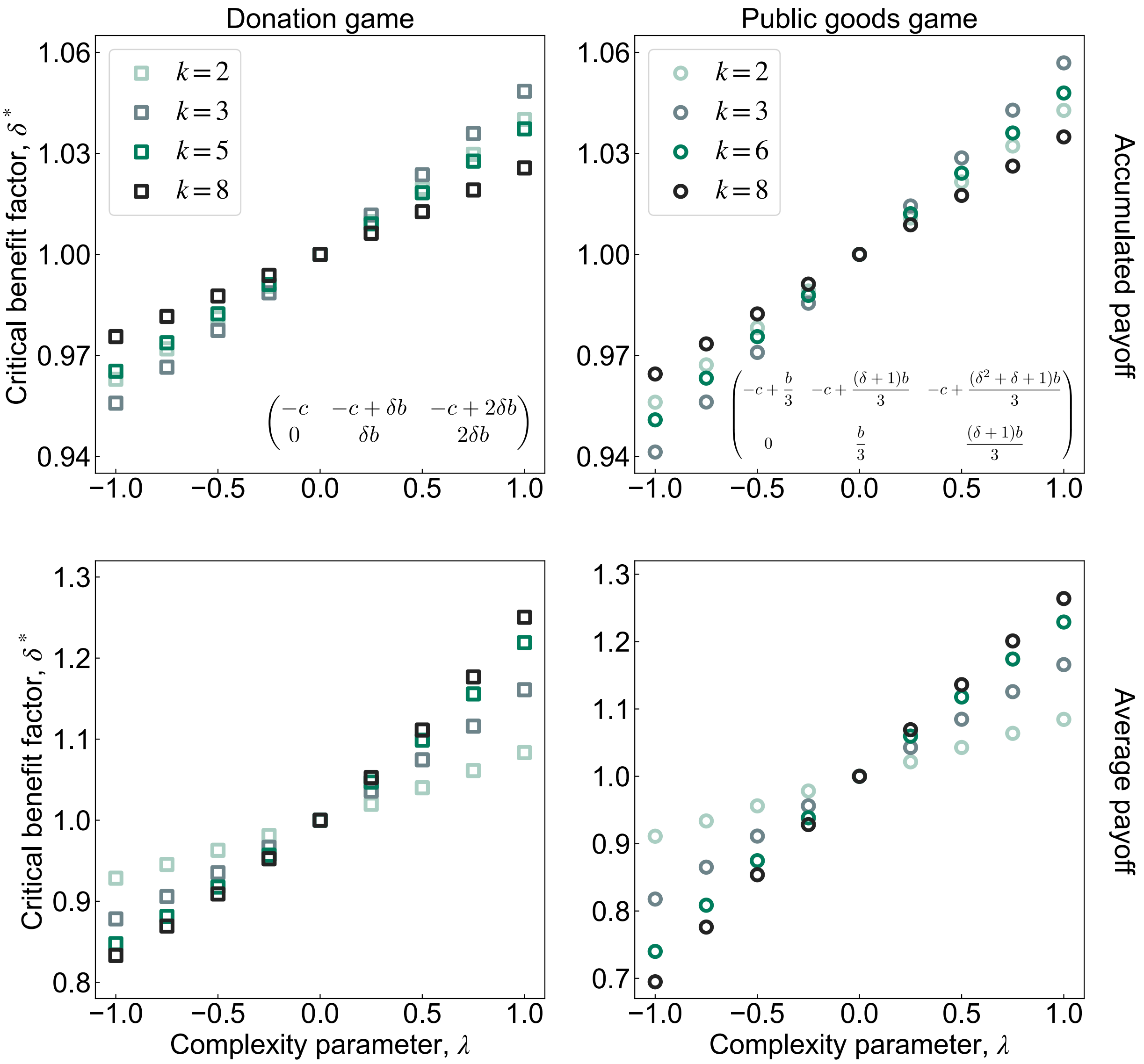


**Supplementary Figure 4: Critical benefit factor for cooperation under complex contagion.** We calculate the critical benefit factor $\delta^*$ that maintains the same cooperation threshold $(b/c)^*$ as in the case without complex contagion ($\lambda = 0$). The critical benefit factor $\delta^*$ increases monotonically with $\lambda$. Similar to the results in Fig. 4c and Supplementary Fig. 3, this trend is more significant under the average-payoff setting.



\end{document}